\documentclass[
reprint,
 amsmath,amssymb,
 aps,
]{revtex4-2}

\usepackage{graphicx}
\usepackage{dcolumn}
\usepackage{bm}
\usepackage{hyperref}
\usepackage{slashed}
\usepackage{subfloat}
\usepackage{pgfplots,subfigure}
\usepackage{booktabs}
\usepackage{orcidlink}
\usepackage{float}
\usepackage{geometry}
\usepackage{array}
\usepackage[table]{xcolor}
\usepackage{booktabs}
\definecolor{acsblue}{RGB}{17,76,139}
\newcolumntype{L}[1]{>{\raggedright\arraybackslash}p{#1}}
\newcolumntype{C}[1]{>{\centering\arraybackslash}p{#1}}

\begin{document}
\fontsize{8}{9}\selectfont
\preprint{APS/123-QED}

\title{Charged particle dynamics in singular spacetimes: hydrogenic mapping and curvature-corrected thermodynamics}

\author{Abdullah Guvendi\orcidlink{0000-0003-0564-9899}}
\email{abdullah.guvendi@erzurum.edu.tr}
\affiliation{Department of Basic Sciences, Erzurum Technical University, 25050, Erzurum, Türkiye}

\author{Semra Gurtas Dogan\orcidlink{0000-0001-7345-3287}}
\email{semragurtasdogan@hakkari.edu.tr (Corresponding Author)}
\affiliation{Department of Medical Imaging Techniques, Hakkari University, 30000, Hakkari, Türkiye}

\author{Omar Mustafa\orcidlink{0000-0001-6664-3859}}
\email{omar.mustafa@emu.edu.tr}
\affiliation{Department of Physics, Eastern Mediterranean University, 99628, G. Magusa, north Cyprus, Mersin 10 - Türkiye}

\author{Hassan Hassanabadi\orcidlink{0000-0001-7487-6898}}
\email{hha1349@gmail.com  }
\affiliation{Department of Physics, Faculty of Science, University of Hradec Králové, Rokitanského 62, 500 03 Hradec Králové, Czechia}
\affiliation{Khazar University, Department of Physics and Electronics, 41 Mahsati Str, AZ1096, Baku, Azerbaijan}

\date{\today}

\begin{abstract}
{\fontsize{8}{9}\selectfont
\setlength{\parindent}{0pt}
We analyze the dynamics of charged test particles in a singular, horizonless spacetime arising as the massless limit of a charged wormhole in the Einstein-Maxwell-Scalar (EMS) framework. The geometry, sustained solely by an electric charge $Q$, features an infinite sequence of curvature singularity shells, with the outermost at $r_\ast = 2|Q|/\pi$ acting as a hard boundary for nonradial motion, while radial trajectories can access it depending on the particle’s charge-to-mass ratio $|q|/m$. Exploiting exact first integrals, we construct the effective potential and obtain circular orbit radii, radial epicyclic frequencies, and azimuthal precession rates. In the weak-field limit ($r\gg |Q|$), the motion reduces to a Coulombic system with small curvature-induced retrograde precession. At large radii, the dynamics maps to a hydrogenic system, with curvature corrections inducing perturbative energy shifts. Approaching $r_\ast$, the potential diverges, producing hard-wall confinement. Curvature corrections also modify the spectral thermodynamics, raising energies and slightly altering entropy and heat capacity. Our results characterize the transition from Newtonian-like orbits to strongly confined, curvature-dominated dynamics.
}
\end{abstract}


\maketitle

\tableofcontents

\section{Introduction}\label{sec:intro}

\setlength{\parindent}{0pt}

Charged spacetimes in general relativity provide fundamental insights into the relationship between electromagnetic fields and spacetime curvature. Classical solutions, such as the Reissner-Nordström (RN) black hole, illustrate how electric charge modifies spacetime geometry, giving rise to inner and outer horizons as well as central singularities~\cite{Reissner1916, Nordström1918, Chandrasekhar1983}. These solutions, however, typically assume the presence of mass. This naturally raises the question: can electric charge alone, in the absence of mass, induce nontrivial spacetime curvature and support physically meaningful structures?

\vspace{0.05cm}

\setlength{\parindent}{0pt}

Wormholes, first introduced by Einstein and Rosen, provide a theoretical framework to explore such questions~\cite{EinsteinRosen1935}. These hypothetical structures connect distant regions of spacetime and, in principle, could act as shortcuts between them. While traversable wormholes generally require exotic matter and often violate classical energy conditions, the inclusion of electric charge adds a new layer of complexity. In charged wormhole geometries \cite{CW-1,CW-2,CW-3,CW-4,CW-5,CW-6}, electromagnetic fields can significantly modify the causal structure and the trajectories of test particles~\cite{Visser1995, MorrisThorne1988}, potentially allowing for configurations that circumvent classical energy condition violations. Recent investigations have extended these considerations to massless configurations, where electric charge alone shapes spacetime curvature. In particular, Turimov \emph{et al.} \cite{Turimov2025} have obtained exact solutions of the EMS field equations for spherically symmetric charged wormholes characterized by mass \(M\) and charge \(Q\). Unlike classical charged black holes, these wormholes reveal a novel mechanism by which charge governs spacetime, motivating a detailed analysis of their dynamics and geometric properties. 

\vspace{0.05cm}

\setlength{\parindent}{0pt}

The spacetime under consideration is described by the static, spherically symmetric metric (in units \(G = c = 1\))~\cite{Turimov2025}:
\begin{equation}
ds^2 = -f(r)\, dt^2 + f(r)^{-1} dr^2 + r^2 \left( d\theta^2 + \sin^2\theta\, d\varphi^2 \right),
\end{equation}
with the metric function:
\begin{equation}
f(r) = \left[ \cosh\!\Big(\frac{\sqrt{M^2 - Q^2}}{r}\Big) + \frac{M}{\sqrt{M^2 - Q^2}} \sinh\!\Big(\frac{\sqrt{M^2 - Q^2}}{r}\Big) \right]^{-2}.
\end{equation}
In the extremal limit \(M \to |Q|\), one has \(\sqrt{M^{2}-Q^{2}}\to 0\). To control this limit, we introduce a small parameter \(\delta\) and set:
\begin{equation}
M = |Q| + \delta, \qquad |\delta| \ll |Q| ,
\end{equation}
which gives:
\begin{equation}
M^{2}-Q^{2} = 2|Q|\,\delta + \mathcal{O}(\delta^{2}),
\qquad
x=\frac{\sqrt{M^{2}-Q^{2}}}{r}
=\frac{\sqrt{2|Q|\delta}}{r}.
\end{equation}
For \(x\ll 1\), the hyperbolic functions admit the standard expansions:
\begin{equation}
\cosh x \simeq 1+\frac{x^{2}}{2},
\qquad
\sinh x \simeq x+\frac{x^{3}}{6}.
\end{equation}
Using these, one finds:
\begin{equation}
\cosh x+\frac{M}{\sqrt{M^{2}-Q^{2}}}\sinh x
\simeq
1+\frac{M}{r}+\mathcal{O}(x^{2})
\longrightarrow
1+\frac{|Q|}{r}
\quad (M\to |Q|).
\end{equation}
Hence, in the extremal limit the metric function reduces to:
\begin{equation}
f(r)\big|_{M\to |Q|}
=
\left(1+\frac{|Q|}{r}\right)^{-2}.
\end{equation}
Introducing the Schwarzschild-like radial coordinate \(R = r + |Q|\), so that \(r = R - |Q|\), the line element becomes:
\begin{equation}
ds^2 = -\left( 1 - \frac{|Q|}{R} \right)^2 dt^2 + \left( 1 - \frac{|Q|}{R} \right)^{-2} dR^2 + (R-|Q|)^2 d\Omega^2,
\end{equation}
where \(d\Omega^2 = d\theta^2 + \sin^2\theta\, d\varphi^2\). This geometry coincides with the extremal RN metric in the radial sector but exhibits a distinct angular sector due to the radial shift \(R \mapsto R - |Q|\). In the neutral limit \(Q \to 0\), it reduces to the classical Papapetrou ``exponential'' wormhole metric~\cite{Papapetrou1943}. For \(|Q| > M\), the hyperbolic functions become trigonometric, yielding oscillatory metrics and generically naked singularities. These features highlight the delicate relationship between mass and charge in determining the global structure of spacetime~\cite{Reissner1916, Nordström1918, Turimov2025}.

\vspace{0.05cm}

\setlength{\parindent}{0pt}

In the massless limit \(M = 0\), electric charge \(|Q|\) alone generates spacetime curvature~\cite{Turimov2025}, resulting in the line element:
\begin{equation}
ds^2 = -\frac{dt^2}{\cos^2(|Q|/r)} + \cos^2(|Q|/r) \left( dr^2 +  r^2d\Omega^2 \right).
\label{o-metric-red}
\end{equation}
This metric exhibits curvature singularities at (see Appendix~A and Figure \ref{fig:curv-inv})
\begin{equation}
r_n = \frac{|Q|}{(n+\tfrac{1}{2})\pi}, \quad n = 0,1,2,\dots,\label{sing-points}
\end{equation}
where \(\cos(|Q|/r)\) vanishes. The explicit calculation of the Ricci scalar \(R(r)\) and the Kretschmann scalar \(K(r) = R_{\alpha\beta\gamma\delta} R^{\alpha\beta\gamma\delta}\) in Appendix~A shows that these scalars diverge exactly at radii \(r_n\). Because these are coordinate-invariant quantities, the shells represent true curvature singularities and act as dynamical barriers confining timelike particles. The spatial distribution and divergence structure of the curvature invariants are directly visualized in Figure~\ref{fig:curv-inv}. Analogies to confined magnetic configurations, such as the Bonnor-Melvin universe \cite{BM,BM-1}, are formal and should not be interpreted as physical equivalence. The accessible radial region between successive singular shells is
\begin{equation}
\frac{|Q|}{(n+\tfrac{3}{2})\pi} < r < \frac{|Q|}{(n+\tfrac{1}{2})\pi}, \quad n = 0,1,2,\dots,
\end{equation}
which can be interpreted classically as a sequence of effective potential wells for large \(n\). The outermost shell (\(n=0\)) is located at:
\begin{equation}
    r_{\ast} = \frac{2|Q|}{\pi},\label{OS}
\end{equation}
which represents an effectively impenetrable boundary for timelike orbits with nonzero angular momentum (\(L \neq 0\)). This property follows from the line element given in~\eqref{o-metric-red}. For purely radial motion (\(L=0\)), the accessibility of \(r_{\ast}\) is governed by the particle's charge-to-mass ratio \(|q|/m\) (see Table \ref{tab:dims}), highlighting the dependence of test particle dynamics on the underlying spacetime geometry. The massless charged configuration examined here is of particular physical interest because it isolates curvature generated purely by electromagnetic fields, providing a minimal and analytically tractable laboratory for studying charge-curvature interactions. Such horizonless, charge-dominated geometries are relevant in the study of exotic compact objects, charged compact remnants, and quantum-gravity-inspired models where horizons may be avoided. This setting allows us to probe how strong electromagnetic fields alone can govern particle dynamics, spectral properties, and accordingly thermodynamic responses in curved spacetime.

\begin{figure*}[!htbp] 
\centering 
\includegraphics[scale=0.65]{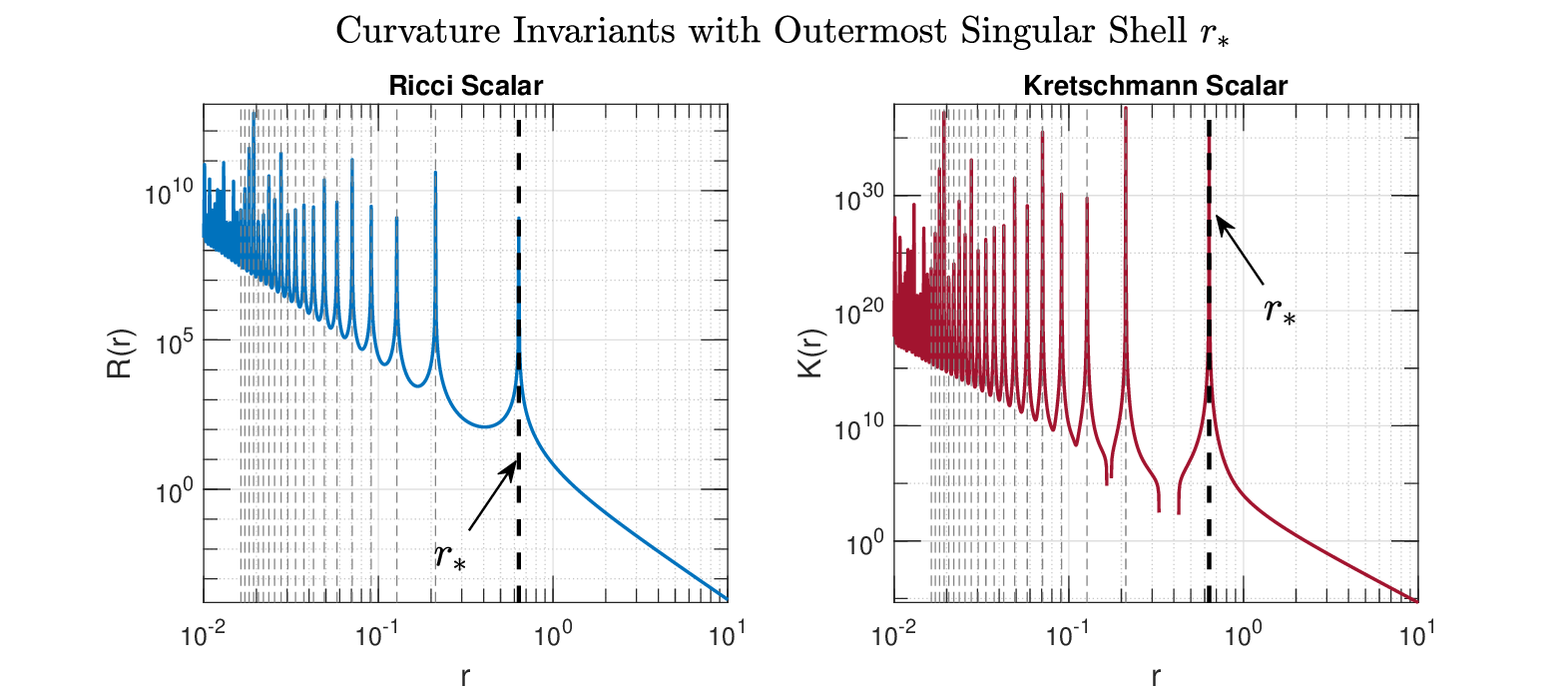} 
\caption{\fontsize{8}{9}\selectfont Curvature invariants (see Appendix A) of the massless spacetime with discrete singular shells. Left: Ricci scalar $R(r)$, Right: Kretschmann scalar $K(r)$, both plotted in log-log scale as functions of radial coordinate $r$. Vertical dashed lines indicate the positions of singular shells $r_n = Q/(\pi/2 + n\pi)$ for $n=0,1,2,3,4...$, with the outermost shell $r_\ast = Q/(\pi/2) \approx 0.6366$ highlighted in black. Both scalars diverge at each singular shell and at the origin, representing true curvature singularities. Away from the singularities, the log-log representation reveals the asymptotic decay $R(r)\sim r^{-4}$ and $K(r)\sim r^{-8}$ for $r \gg r_\ast$, highlighting the fall-off of curvature in the weak-field region. Values used: $Q=1$, radial range $r \in [0.01\,Q,10\,Q]$. The log-log scale emphasizes the power-law divergence near each singular shell and the central singularity, as well as the asymptotic behavior far from the shells. The outermost shell $r_\ast$ acts as the first hypersurface of infinite curvature, demonstrating the nontrivial geometric structure of this massless spacetime. The sequence of singular shells constrains geodesic motion and defines regions of extreme tidal forces.}
\label{fig:curv-inv} 
\end{figure*}

\vspace{0.05cm}

\setlength{\parindent}{0pt}

In the far-field regime (\(r \gg |Q|\)) or for weak charge (\(|Q| \ll r\)), the metric functions expand as:
\begin{align}
\cos^{-2}\!\Big(\frac{|Q|}{r}\Big) &= 1 + \Big(\frac{|Q|}{r}\Big)^{2} + O\!\Big(\frac{|Q|^{4}}{r^{4}}\Big),
\nonumber\\
\cos^{2}\!\Big(\frac{|Q|}{r}\Big) &= 1 - \Big(\frac{|Q|}{r}\Big)^{2} + O\!\Big(\frac{|Q|^{4}}{r^{4}}\Big),
\end{align}
showing that the spacetime is asymptotically Minkowskian, with curvature corrections decaying as \((|Q|/r)^2\). Thus, the geometry is regular at large distances, whereas its short-distance structure is entirely governed by the electric charge, highlighting the nontrivial role of charge in the absence of mass. It is important to emphasize that the Israel junction conditions \cite{ref-1,ref-2} apply for $r \neq r_n$, in which case the surface energy density and pressure are given, respectively, by:
\[
\sigma(r) = -\frac{1}{2\pi r \cos\!\left(\frac{|Q|}{r}\right)}, 
\qquad 
P(r) = \frac{r - |Q| \tan\!\left(\frac{|Q|}{r}\right)}{8\pi r^2 \cos\!\left(\frac{|Q|}{r}\right)}.
\]
At $r = r_n$, however, the Israel junction conditions cease to be applicable, since each hypersurface represents a genuine curvature singularity rather than a thin shell.

\vspace{0.05cm}
\setlength{\parindent}{0pt}

The motion of charged test particles \cite{TP-1,TP-2,TP-3,TP-4,TP-4,TP-5,TP-6} is governed by the Lorentz force in curved spacetime~\cite{L-1,L-2,L-3,LL-3,L-4,CLF} (see also \cite{2-1,2-2,2-3,2-4,2-5,TP-2,2-7,2-8,guvendi,hassan,hassan-2}). In this work, we focus exclusively on timelike trajectories. The singular shell structure may give rise to a rich variety of dynamics, including bounded motion, scattering, and capture, all constrained by the outermost shell \(r_{\ast}\). A detailed effective potential analysis reveals how the singular shells regulate orbital motion and determine the stability of circular orbits. Remarkably, weak-field circular orbits may exhibit retrograde precession \cite{retro}, opposite in sign to the prograde advance \cite{prog,O-P} observed in Schwarzschild and RN spacetimes, providing a clear dynamical signature of the charge-induced geometry~\eqref{o-metric-red}. More broadly, such charge-dominated spacetimes offer a unique framework for studying exotic objects, semiclassical instabilities, and naked singularities, with implications for testing deviations from general relativity in extreme regimes.

\vspace{0.05cm}
\setlength{\parindent}{0pt}

In this paper, we present a comprehensive investigation of the dynamics of massive, charged test particles in the massless, charge-induced geometry~\eqref{o-metric-red}. We analyze the effective potentials, stability criteria, and the influence of curvature singularity shells, obtaining analytical solutions and deriving weak-field approximations to connect with Newtonian intuition. In the weak-field regime, the system is semiclassically mapped to a hydrogenic model, where curvature-induced corrections yield controlled perturbative energy shifts. The study is further extended to spectral thermodynamics, demonstrating how these curvature effects systematically modify free and internal energies, as well as entropy and heat capacity. The paper is organized as follows: Section~\ref{sec:dynamics} introduces the spacetime geometry and associated electromagnetic field, adopting the gauge \(A_t = -\tan(|Q|/r)\), which reduces to the Coulomb potential \(A_t \simeq -|Q|/r\) in the weak-field limit, along with the equations of motion and construction of the effective potential. Section~\ref{sec:analysis} presents a detailed study of orbital dynamics and stability. Section~\ref{sec:mapping} maps the system to a one-electron atom, highlighting similarities, perturbative curvature corrections, and limits of validity. Section~\ref{sec:thermo} extends the analysis to spectral thermodynamics, illustrating the impact of curvature corrections on thermodynamic properties. Finally, Section~\ref{sec:discussions} summarizes the main results and outlines potential directions for future research.

\section{Charged Particle Dynamics}\label{sec:dynamics}

\setlength{\parindent}{0pt}

\begin{figure}[!htbp] 
\centering 
\includegraphics[scale=0.55]{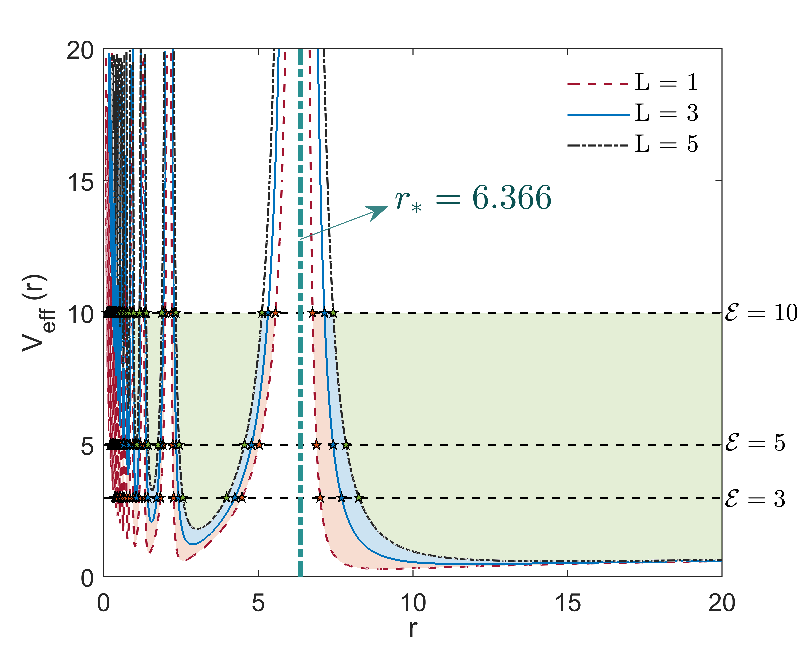} 
\caption{\fontsize{8}{9}\selectfont Effective radial potentials $V_{\rm eff}(r)$ for different angular momentum states $L = 1, 3, 5$ of a particle with charge $q=-1$ and unit mass $m=1$ in the presence of a singular shell located at \(r_* = 2|Q|/\pi\) ($Q = 10$). Colored curves represent the effective potentials for each $L$ state, with the corresponding colored stars indicating the classical turning points for energies $\mathcal{E} = 3, 5, 10$. Shaded regions highlight the classically allowed radial motion ($\mathcal{E} > V_{\rm eff}(r)$), and the dashed vertical line marks the outermost singular shell position $r_{\ast}$. This visualization illustrates how the effective potential and the allowed regions depend on angular momentum and energy levels (see Table \ref{tab:dims}).}
\label{fig:pott} 
\end{figure}

We consider the motion of a charged test particle with mass \(m\) and charge \(q\) in the curved spacetime geometry \eqref{o-metric-red}, introduced in Section~\ref{sec:intro}. The dynamics of the particle are determined by the Lagrangian~\cite{Lag,L-5,EM-Lag}
\begin{equation}
\mathcal{L} = \frac{m}{2}\, g_{\mu\nu}\,\dot{x}^\mu \dot{x}^\nu + q\, A_\mu \dot{x}^\mu,
\label{Eq:Lagrangian}
\end{equation}
where the dot denotes differentiation with respect to the proper time \(\tau\), i.e. \(\dot{x}^\mu \equiv dx^\mu/d\tau\). The first term represents the kinetic contribution associated with motion in the curved background geometry, while the second term implements the minimal coupling to the electromagnetic four-potential \(A_\mu\) (see Table \ref{tab:dims}). The equations of motion are obtained by applying the Euler-Lagrange equations to the Lagrangian~\eqref{Eq:Lagrangian}, namely:
\begin{equation}
\frac{d}{d\tau}\!\left(\frac{\partial \mathcal{L}}{\partial \dot{x}^\mu}\right) - \frac{\partial \mathcal{L}}{\partial x^\mu} = 0.
\end{equation}
Evaluating the first term, we find the canonical momentum:
\begin{equation}
\frac{\partial \mathcal{L}}{\partial \dot{x}^\mu} = m\, g_{\mu\nu}\,\dot{x}^\nu + q\, A_\mu,
\end{equation}
and differentiating with respect to proper time yields:
\begin{equation}
\frac{d}{d\tau}\!\left(\frac{\partial \mathcal{L}}{\partial \dot{x}^\mu}\right) 
= m\Big(\partial_\lambda g_{\mu\nu}\,\dot{x}^\lambda \dot{x}^\nu + g_{\mu\nu}\,\ddot{x}^\nu\Big)
+ q\,\partial_\lambda A_\mu\,\dot{x}^\lambda.
\end{equation}
On the other hand, the explicit coordinate dependence of the Lagrangian contributes:
\begin{equation}
\frac{\partial \mathcal{L}}{\partial x^\mu} 
= \frac{m}{2}\,\partial_\mu g_{\alpha\beta}\,\dot{x}^\alpha \dot{x}^\beta + q\,\partial_\mu A_\alpha\,\dot{x}^\alpha.
\end{equation}
Substituting these expressions into the Euler-Lagrange equation leads to:
\begin{equation}
m\Big(g_{\mu\nu}\ddot{x}^\nu + \partial_\lambda g_{\mu\nu}\,\dot{x}^\lambda \dot{x}^\nu - \tfrac{1}{2}\,\partial_\mu g_{\alpha\beta}\,\dot{x}^\alpha \dot{x}^\beta\Big) 
= q\big(\partial_\mu A_\nu - \partial_\nu A_\mu\big)\dot{x}^\nu.
\end{equation}
At this stage it is natural to recognize the antisymmetric electromagnetic field strength tensor:
\begin{equation}
F_{\mu\nu} = \partial_\mu A_\nu - \partial_\nu A_\mu,
\end{equation}
which allows the right-hand side to be written in compact form. By raising an index with the inverse metric and noting that the combination of metric derivatives reproduces the Christoffel symbols of the Levi-Civita connection,
\begin{equation}
\Gamma^\sigma_{\alpha\beta} = \tfrac{1}{2}\, g^{\sigma\mu}\big(\partial_\alpha g_{\mu\beta} + \partial_\beta g_{\mu\alpha} - \partial_\mu g_{\alpha\beta}\big),
\end{equation}
the equation of motion assumes the form:
\begin{equation}
m\Big(\ddot{x}^\sigma + \Gamma^\sigma_{\alpha\beta}\,\dot{x}^\alpha \dot{x}^\beta\Big) 
= q\,F^\sigma{}_{\nu}\,\dot{x}^\nu.
\end{equation}
This result can be expressed even more transparently in covariant notation. Writing \(\nabla_{\dot{x}}\) for the covariant derivative along the worldline, one obtains \cite{CLF}:
\begin{equation}
m\,\nabla_{\dot{x}} \dot{x}^\mu = q\, F^\mu{}_\nu \,\dot{x}^\nu,
\end{equation}
which is the covariant Lorentz force law describing the trajectory of a charged particle subject simultaneously to gravitational and electromagnetic fields. Here, \(F_{\mu\nu}\) encodes the electromagnetic field, while the gravitational influence enters through the connection \(\Gamma^\sigma_{\alpha\beta}\). Throughout this work, we adopt units with \(G = c = 1\), unless stated otherwise, and employ the gauge choice corresponding to the \(M \to 0\) limit of Eq.~(17) in \cite{Turimov2025}):
\begin{equation}
A_t = -\tan\!\left(\frac{|Q|}{r}\right), \label{EM-field}
\end{equation}
which asymptotically reduces to the Coulomb form \(A_t \to -|Q|/r\) as \(r \to \infty\), thereby ensuring the correct flat-space limit. Exploiting spherical symmetry, we restrict motion to the equatorial plane \(\theta=\pi/2\) \cite{Review}. In this plane, the Lagrangian \eqref{Eq:Lagrangian} simplifies to:
\begin{equation}
\begin{split}
\mathcal{L} &= \frac{m}{2} \Biggl[-\frac{\dot{t}^{\,2}}{\cos^2(|Q|/r)} 
+ \cos^2(|Q|/r)\bigl(\dot{r}^{\,2} + r^2 \dot{\varphi}^{\,2}\bigr)\Biggr] 
- q\, \tan(|Q|/r)\, \dot{t}.
\label{Lagrangian-2D}
\end{split}
\end{equation}
The existence of timelike and rotational Killing vectors ensures two conserved quantities: the energy \(\mathcal{E}\) and angular momentum \(L\)~\cite{PLB-2025}. These arise from the canonical momenta:
\begin{align}
p_t &= \frac{\partial \mathcal{L}}{\partial \dot{t}} 
= -\frac{m}{\cos^2(|Q|/r)} \dot{t} - q\,\tan(|Q|/r) 
\equiv -\mathcal{E}, \nonumber \\ 
p_\varphi &= \frac{\partial \mathcal{L}}{\partial \dot{\varphi}} 
= m r^2 \cos^2(|Q|/r)\, \dot{\varphi} 
\equiv L.
\label{Can-momenta}
\end{align}
Solving for the velocities yields:
\begin{equation}
\dot{t} = \frac{\mathcal{E} - q\, \tan(|Q|/r)}{m}\, \cos^2(|Q|/r), 
\qquad
\dot{\varphi} = \frac{L}{m r^2 \cos^2(|Q|/r)}.
\label{Coor-velocity}
\end{equation}
Substituting into the timelike condition \(g_{\mu\nu}\dot{x}^\mu \dot{x}^\nu=-1\) gives:
\[
m^2\dot{r}^2 = (\mathcal{E}-q \tan(|Q|/r))^2 - \frac{m^2}{\cos^2(|Q|/r)} - \frac{L^2}{ r^2 \cos^4(|Q|/r)}.
\]
Defining the energy branches \cite{E-branches}:
\begin{equation}
\mathcal{E}_\pm(r) \equiv q\,\tan\!\left(\frac{|Q|}{r}\right) 
\pm \sqrt{\frac{m^{2}}{\cos^{2}(|Q|/r)} + \frac{L^{2}}{r^{2} \cos^{4}(|Q|/r)}}\,.
\label{E-branches}
\end{equation}
The effective potential per unit mass, shown in Figure~\ref{fig:pott}, is defined as \cite{E-branches}:
\begin{equation}
V_{\text{eff}}(r) = \frac{\mathcal{E}_{+}(r)}{m}.
\end{equation}
Accordingly, the binding energy is
\begin{equation}
\mathcal{E}_\text{bind}(r) \equiv V_{\rm eff}(r) - 1.
\end{equation}
This definition follows from considering the particle at rest at infinity: in this limit, \(\mathcal{E}_+ \to m\) and \(V_{\rm eff}\to 1\), so \(\mathcal{E}_\text{bind} \to 0\). Regions with \(V_{\rm eff}<1\) correspond to bound motion, while \(V_{\rm eff}>1\) indicates unbound motion. The radial motion is allowed where \(\mathcal{E} \ge \mathcal{E}_+\), linking turning points directly to the effective potential \cite{E-branches}. Factorizing the radial equation:
\begin{equation}
m^2 \dot{r}^2 = (\mathcal{E} - \mathcal{E}_+)(\mathcal{E} - \mathcal{E}_-) \equiv \mathcal{R}(r),
\end{equation}
makes clear that \(\dot{r}^2 \ge 0\) only in classically allowed regions. Circular orbits occur at \(r_c>r_{\ast}\) where \(\mathcal{E}_+'(r_c)=0\), with stability determined via proper-time radial epicyclic frequency \cite{epic-fre}:
\begin{equation}
\omega_r^2 \equiv -\frac{\mathcal{R}''(r_c)}{2 m^2} = \frac{\mathcal{E}_+''(r_c) \bigl[\mathcal{E}_+(r_c) - \mathcal{E}_-(r_c)\bigr]}{2 m^2}.
\label{omega_exact}
\end{equation}
In the weak-field limit, \(\mathcal{E}_+-\mathcal{E}_- \simeq 2m\), giving \(\omega_r^2 \simeq \mathcal{E}_+''(r_c)/m\). Stability requires \(\omega_r^2>0\) or \(V_{\rm eff}''(r_c)>0\). The coordinate-time radial frequency is
\begin{equation}
\dot{t}\big|_{r_c} = \frac{\mathcal{E}-q\tan(|Q|/r_c)}{m}\cos^2(|Q|/r_c), 
\qquad \Omega_r = \frac{\omega_r}{\dot{t}\big|_{r_c}}.
\end{equation}

\vspace{0.05cm}
\setlength{\parindent}{0pt}

Figure~\ref{fig:pott} illustrates how \(V_{\rm eff}(r)\) depends on \(L\) and \(Q\). Key features include: (i) higher \(L\) increases the centrifugal barrier, moving circular orbits outward; (ii) the depth of \(V_{\rm eff}\) indicates the strength of binding, with lower \(V_{\rm eff}\) corresponding to more tightly bound orbits; (iii) the combined effect of spacetime curvature and the electric field produces barriers absent in RN spacetimes, making \(r_{\ast}\) impenetrable for \(L\neq0\); (iv) for radial motion \((L=0)\), accessibility of \(r_{\ast}\) depends on \(|q|/m\) and \(\mathcal{E}\). This figure thus encapsulates turning points, classically allowed regions, and the influence of conserved quantities on orbital stability.

\section{Analysis of Radial Motion, Particle Orbits, and Stability}
\label{sec:analysis}

\setlength{\parindent}{0pt}

We now analyze in detail the dynamics of a classical charged test particle with rest mass \(m\) and charge \(q\) in the background geometry \eqref{o-metric-red}. Owing to the stationarity and spherical symmetry of the spacetime, there exist two Killing vectors, \(\partial_t\) and \(\partial_\varphi\), which yield conserved energy and angular momentum along the particle's worldline. These constants of motion reduce the problem to an effective one-dimensional radial equation without the need for weak-field approximations \cite{Review}.

\subsection{Radial Motion and Effective Potential}

\setlength{\parindent}{0pt}

As shown in Sec.~\ref{sec:dynamics}, the radial dynamics can be cast in terms of two energy branches \(\mathcal{E}_\pm(r)\), associated with future- and past-directed timelike trajectories. Classical motion occurs when \(\mathcal{E} \ge \mathcal{E}_+(r)\) for future-directed trajectories, and \(\mathcal{E} \le \mathcal{E}_-(r)\) for past-directed trajectories. The spacetime singularities occur at the discrete radii determined by \(\cos(|Q|/r)=0\) (cf. Eq.~\eqref{sing-points}), where the effective energies \(|\mathcal{E}_\pm|\) diverge. These singular hypersurfaces act as absolute kinematic barriers. The outermost such barrier, located at \(r_{\ast} = 2|Q|/\pi\), bounds all physically realizable trajectories. For purely radial motion (\(L=0\)), the divergences of the terms \([\mathcal{E}-q\,\tan\,(|Q|/r)]^2\) and \(m^2 \sec^2\,(|Q|/r)\) both become relevant as \(r \to r_{\ast}\). Now, let us introduce the dimensionless variable \(u = |Q|/r\), mapping spatial infinity (\(r \to \infty\)) to \(u \to 0\) and the singular barrier (\(r = r_{\ast}\)) to \(u \to \pi/2\). Since \(\tan u \sim \sec u\) as \(u \to \pi/2\), the near-barrier behavior depends sensitively on the ratio \(|q|/m\) and the conserved canonical energy \(\mathcal{E}\). In particular, for \(|q|/m \lesssim 1\), the particle is repelled before reaching \(r_{\ast}\), while for \(|q|/m \gtrsim 1\), the electrostatic attraction may partially compensate, allowing closer approach. To systematically analyze radial motion, we define the radial function
\begin{equation}
\mathcal{R}(r) \equiv \bigl[\mathcal{E} - q\,\tan\, (|Q|/r)\bigr]^2 
- \frac{m^2}{\cos^2\,(|Q|/r)} 
- \frac{L^2}{r^2 \cos^4\,(|Q|/r)},
\label{eq:Rdef-extended}
\end{equation}
so that the radial equation reduces to
\begin{equation}
m^2 \dot{r}^2 = \mathcal{R}(r), \qquad \mathcal{R}(r) \geq 0.\label{rad:eq}
\end{equation}
Physically, \(\mathcal{R}(r)\) plays the role of the “radial kinetic energy squared”: the particle can move only where \(\mathcal{R}(r)\ge 0\) (see Table \ref{tab:dims}). Turning points occur at \(\mathcal{R}(r)=0\), corresponding to \(V_{\rm eff}(r)=\mathcal{E}_{+}/m\). For nonzero angular momentum, the centrifugal term \(\sim L^2/(r^2 \cos^4\,(|Q|/r))\) diverges at \(r_{\ast}\), preventing penetration. Hence the physical domain is \(r>r_{\ast}\). For orbits with \(L\neq0\), circular orbits at \(r=r_c\) satisfy simultaneously \cite{Review}
\begin{equation}
\mathcal{R}(r_c) = 0, \qquad \mathcal{R}'(r_c) = 0.
\end{equation}
The radial acceleration can be written as
\begin{equation}
m^2 \ddot{r} = \frac{1}{2} \mathcal{R}'(r).\label{rad-eq}
\end{equation}

\subsection*{Stability of Circular Orbits}

\setlength{\parindent}{0pt}

To study stability, let us consider a small radial perturbation around a circular orbit \cite{HO}:
\begin{equation}
r(t) = r_c + \delta r(t), \qquad |\delta r| \ll r_c,\label{rad-pert}
\end{equation}
and linearize the radial equation. Since \(\mathcal{R}'(r_c)=0\) for a circular orbit, we have to leading order:
\begin{equation}
\mathcal{R}'(r) \approx \mathcal{R}''(r_c)\,\delta r,
\end{equation}
and substitution into the radial acceleration \(m^2 \ddot r = \tfrac{1}{2}\mathcal{R}'(r)\) (cf. Eq.~\eqref{rad-eq}) yields the harmonic-oscillator equation:
\begin{equation}
m^2 \ddot{\delta r} = \frac{1}{2}\mathcal{R}''(r_c)\,\delta r
\quad\Longrightarrow\quad
\ddot{\delta r} = \frac{\mathcal{R}''(r_c)}{2 m^2}\,\delta r.
\end{equation}
Defining the proper-time radial epicyclic frequency \(\omega_r\) with the conventional sign convention for stability,
\begin{equation}
\omega_r^2 \equiv -\,\frac{\mathcal{R}''(r_c)}{2 m^2},
\end{equation}
so that \(\omega_r^2>0\) corresponds to harmonic (stable) oscillations and \(\omega_r^2<0\) to exponential instability. To make the connection between \(\mathcal{R}''(r_c)\) and the energy branches explicit, recall the factorized form of the radial function:
\begin{equation}
\mathcal{R}(r) = \bigl[\mathcal{E}-\mathcal{E}_+(r)\bigr]\bigl[\mathcal{E}-\mathcal{E}_-(r)\bigr],
\end{equation}
which follows directly from the definition \(\mathcal{R}= (\mathcal{E}-\mathcal{E}_+)(\mathcal{E}-\mathcal{E}_-)\). Differentiating twice and evaluating at the circular orbit condition \(\mathcal{E}=\mathcal{E}_+(r_c)\) gives:
\begin{equation}
\mathcal{R}''(r_c) 
= -\bigl[\mathcal{E}_+(r_c)-\mathcal{E}_-(r_c)\bigr]\,\mathcal{E}_+''(r_c),
\end{equation}
so that
\begin{equation}
\omega_r^2 \;=\; \frac{\mathcal{E}_+''(r_c)\,\bigl[\mathcal{E}_+(r_c)-\mathcal{E}_-(r_c)\bigr]}{2 m^2}.
\label{omega_Epm}
\end{equation}
In the weak-field, nonrelativistic limit one has \(\mathcal{E}_+(r_c)-\mathcal{E}_-(r_c)\simeq 2m\), and hence:
\begin{equation}
\omega_r^2 \simeq \frac{\mathcal{E}_+''(r_c)}{m},
\end{equation}
recovering the familiar relation between the curvature of the effective potential and the epicyclic frequency. Thus stability is equivalent to a local minimum of the effective potential \(V_{\rm eff}(r)=\mathcal{E}_+(r)/m\). The observable (coordinate-time) radial frequency \(\Omega_r\) is obtained from the proper-time frequency by the redshift factor \(d\tau/dt\) evaluated on the circular orbit. Using \(\dot t = [\mathcal{E}-q\tan(|Q|/r)]\cos^2(|Q|/r)/m\) (cf. Eq.~\eqref{Coor-velocity}), one finds:
\begin{equation}
\Omega_r = \omega_r \left(\frac{d\tau}{dt}\right)_{r_c}, \qquad 
\frac{d\tau}{dt}\Big|_{r_c} = \frac{m}{\bigl[\mathcal{E}-q\tan(|Q|/r_c)\bigr]\cos^2(|Q|/r_c)}.
\end{equation}
This relation makes explicit how the combination of gravitational redshift (encoded in \(\cos^2(|Q|/r_c)\)) and the electromagnetic gauge potential (encoded in \(\mathcal{E}-q\tan(|Q|/r_c)\)) suppresses or enhances the coordinate-time oscillation frequency relative to the proper-time frequency. The combined effect of angular momentum, charge-to-mass ratio, and the singular barrier \(r_{\ast}\) governs both the allowed radial domain and the stability properties of circular orbits: the centrifugal term produces an increasingly steep potential near \(r_{\ast}\) (excluding penetration for \(L\neq0\)), while the electrostatic interaction and mass term set the detailed curvature of \(V_{\rm eff}(r)\) that determines \(\mathcal{E}_+''(r_c)\) and hence \(\omega_r^2\).

\subsection{Weak-Field Approximation and Orbital Stability}\label{weak-field}

\setlength{\parindent}{0pt}

\begin{figure*}[ht]
\centering
\includegraphics[scale=0.60]{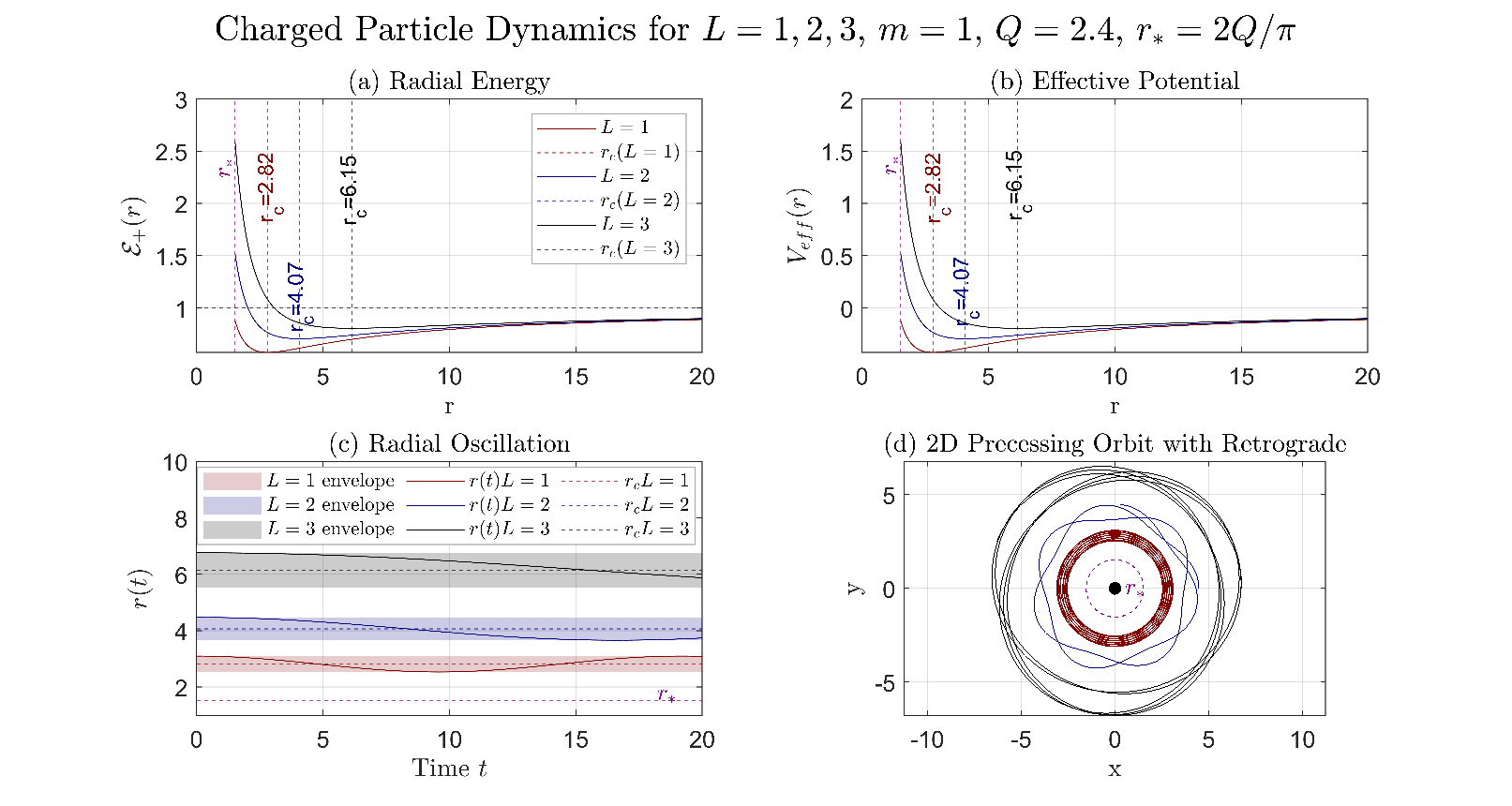}
\caption{\fontsize{8}{9}\selectfont Dynamics of a charged particle with unit mass $m = 1$ and charge $q = -1$ around a central Coulomb charge $|Q| = 2.4$ for angular momenta $L = 1, 2, 3$. (\textbf{a}) Radial energy $\mathcal{E}_+(r)$ showing contributions from the rest mass (black dashed line), Coulomb interaction, and angular momentum, with circular orbits $r_c$ indicated by vertical dashed lines and the outermost barrier $r_{\ast}$ highlighted in purple. (\textbf{b}) Effective potential $V_{\rm eff}(r)$ illustrating the radial dependence of the potential energy, with circular orbits and $r_{\ast}$ similarly marked (purple). (\textbf{c}) Radial oscillations $r(t)$ around circular orbits with shaded envelopes representing the oscillation amplitude, and $r_{\ast}$ shown as a purple dashed line. (\textbf{d}) Two-dimensional precessing orbits in the $xy$ plane, exhibiting retrograde precession around the central charge (black dot), with maximum and minimum radial excursions, and the outermost barrier $r_{\ast}$ shown as a dashed purple circle (see also Table \ref{tab:dims}).}\label{fig:pot-1}
\end{figure*}

\vspace{0.05cm}
\setlength{\parindent}{0pt}

In the weak-field regime, defined by radial distances much larger than the characteristic scale of the central charge (note that $r$ in units of $Q$), $r \gg |Q|$, the spacetime metric approaches the Minkowski form, with small perturbations due to both the electromagnetic field of the central charge and the curvature it induces. In this limit, the dynamics of a charged particle can be described by an effective energy function $\mathcal{E}_+(r)$, which includes contributions from the particle's rest mass, electromagnetic potential, orbital angular momentum, and leading curvature corrections. Expanding $\mathcal{E}_+(r)$ in powers of $1/r$ up to second order gives:  
\begin{equation}
\begin{split}
\mathcal{E}_+(r) &\simeq m + \frac{q |Q|}{r} + \frac{L^2}{2 m r^2} + \frac{m Q^2}{2 r^2} + \mathcal{O}(r^{-3}),\\
\Rightarrow \mathcal{E}_+(r) - m &\simeq \frac{q |Q|}{r} + \frac{L^2}{2 m r^2} + \frac{m Q^2}{2 r^2} = -\frac{\kappa}{r} + \frac{\beta}{r^2},\label{E-expansion}
\end{split}
\end{equation}
where we define $\kappa = |q Q|$, $\beta = (L^2 + m^2 Q^2)/(2 m)$ and $q<0$. In this decomposition, the first term represents the attractive Coulomb interaction between the particle and the central charge. The second term corresponds to the centrifugal barrier arising from the orbital angular momentum, which prevents the particle from collapsing into the central charge. The third term represents the leading-order correction due to spacetime curvature induced by the central charge, which slightly modifies the effective potential at large distances. Terms of order $\mathcal{O}(r^{-3})$ and higher are negligible in this approximation and do not significantly influence the orbital motion in the weak-field regime. A circular orbit corresponds to a radius $r_c$ where the radial derivative of the effective energy vanishes. Physically, this condition reflects the balance between the attractive and repulsive contributions to the radial force acting on the particle. Differentiating the energy function with respect to $r$ yields:  
\begin{equation}
\mathcal{E}_+'(r) = - \frac{q Q}{r^2} - \frac{L^2}{m r^3} - \frac{m Q^2}{r^3} + \mathcal{O}(r^{-4}).
\end{equation}
At leading order, we can neglect the curvature term proportional to $Q^2 / r^3$, since it is subdominant at large radii. This reduces the circular orbit condition to the classical balance between the Coulomb force and the centrifugal barrier:  
\begin{equation}
\frac{L^2}{m r_c^3} = - \frac{q Q}{r_c^2}, \quad q Q < 0 \quad \Rightarrow \quad r_c = \frac{L^2}{m |q Q|}.
\label{COC}
\end{equation}
Here, the restriction $q Q < 0$ ensures that the Coulomb interaction is attractive, allowing for stable circular orbits. Including the curvature correction to next-to-leading order slightly increases the circular orbit radius:  
\begin{equation}
r_c \simeq \frac{1}{|q Q|} \left( \frac{L^2}{m} + m Q^2 \right),
\end{equation}
which reduces to the leading-order expression when $Q^2 / r_c^2 \ll 1$. This demonstrates that the curvature of spacetime effectively contributes a small repulsive term, increasing the orbital radius for a given angular momentum. Physically, this reflects the fact that curvature-induced modifications to the potential slightly oppose the central Coulomb attraction. The stability of circular orbits is characterized by the radial epicyclic frequency, which describes the particle’s small oscillations around the circular orbit. A positive radial frequency indicates stable oscillations, while a negative or imaginary frequency would signal instability. The radial epicyclic frequency is defined as:  
\begin{equation}
\omega_r^2 \simeq \frac{1}{m} \mathcal{E}_+''(r_c),
\end{equation}
with the second derivative of the effective energy given by:  
\begin{equation}
\mathcal{E}_+''(r) = \frac{2 q Q}{r^3} + \frac{3 L^2}{m r^4} + \frac{3 m Q^2}{r^4} + \mathcal{O}(r^{-5}).
\end{equation}
Evaluating this at the circular orbit radius $r_c$ using \eqref{COC}, the leading-order term yields:  
\begin{equation}
\mathcal{E}_+''(r_c) \simeq \frac{|q Q|}{r_c^3} \left[ 1 + \mathcal{O}\left( \frac{m Q^2}{L^2} \right) \right] > 0,
\end{equation}
confirming the stability of the orbit under small radial perturbations. Consequently, the proper-time radial epicyclic frequency can be expressed as:  
\begin{equation}
\omega_r \simeq \frac{m |q Q|^2}{L^3},
\end{equation}
up to minor corrections (cancelled) from the curvature term. This relation has a clear physical interpretation: stronger Coulomb attraction increases the radial oscillation frequency, while larger angular momentum reduces it due to the broader orbits associated with higher $L$. In the limit $m \to 0$, the radial frequency vanishes, consistent with the absence of a restoring force for massless particles. To investigate the azimuthal motion and the associated orbital precession \cite{O-P}, it is convenient to define an effective central potential incorporating both Coulomb and curvature effects:  
\begin{equation}
U(r) = \frac{q Q}{r} + \frac{m Q^2}{2 r^2}.
\end{equation}
The circular orbit condition can be equivalently written as $L^2 = m\, r^3\, U'(r)$, and the proper-time frequencies for small radial and azimuthal oscillations are given by: 
\begin{equation}
\omega_\varphi^2 = \frac{1}{m r} U'(r), \qquad \omega_r^2 = \frac{1}{m} \left( U''(r) + \frac{3 L^2}{m r^4} \right). 
\end{equation}
Differentiating the potential provides:
\begin{equation}
U'(r) = -\frac{q Q}{r^2} - \frac{m Q^2}{r^3}, \qquad U''(r) = \frac{2 q Q}{r^3} + \frac{3 m Q^2}{r^4}.
\end{equation}
Substituting these into the frequency expressions shows that the radial epicyclic frequency is dominated by the Coulomb term, while the azimuthal frequency is slightly reduced due to the curvature contribution:  
\begin{equation}
\omega_\varphi^2 \simeq - \frac{q Q}{m r^3} - \frac{Q^2}{r^4}, \qquad \omega_r^2 \simeq - \frac{q Q}{m r^3}.
\end{equation}
This difference in frequencies gives rise to a retrograde precession, meaning that the orbit slowly rotates backward relative to the radial oscillations. The precession per orbit can be expressed as:  
\begin{equation}
\begin{split}
\Delta \varphi &\simeq 2 \pi \left( 1 - \frac{\omega_\varphi}{\omega_r} \right) \simeq 2 \pi \left( 1 - \sqrt{ 1 + \frac{m Q^2}{|q Q| r_c} } \right)\\
&\simeq - \frac{\pi m Q^2}{|q Q| r_c} = - \frac{\pi m^2 Q^2}{L^2}.   
\end{split}
\end{equation}
The negative sign explicitly confirms that the precession is retrograde \cite{retro}. Its magnitude is small, consistent with the weak-field approximation, and scales as $Q^2 / L^2$, indicating that curvature effects become significant only for tight orbits or large central charges. Thus, the weak-field approximation provides a clear and physically intuitive description of orbital dynamics in the presence of a central charged source. Circular orbits exist and are stable under small radial perturbations. Radial oscillation frequencies increase with stronger Coulomb attraction and decrease with higher angular momentum. The curvature-induced modification of the azimuthal frequency leads to a small retrograde precession, generalizing classical Keplerian dynamics to include leading-order corrections. The effective potential $U(r)$ offers a concise framework to understand how electromagnetic forces, centrifugal barriers, and spacetime curvature together determine the orbital structure of charged particles.

\begin{table*}[htbp]
\centering
\fontsize{8}{9}\selectfont
\renewcommand{\arraystretch}{1.4}

\begin{tabular}{l c l}
\hline
\textbf{Quantity} 
& \textbf{Definition / Representative Expression} 
& \qquad \textbf{Dimension} \\
\hline

Invariant spacetime interval 
& $ds^{2} = g_{\mu\nu}\,dx^{\mu}dx^{\nu}$ 
& \qquad $[ds^{2}] = L^{2}$ \\

Spacetime coordinates and proper time 
& $x^\mu=(t,r,\varphi),\;\tau$ 
& \qquad $[x^\mu]=[\tau]=L$ \\

Test-particle rest mass 
& $m$ 
& \qquad $[m] = L$ \\

Conserved energy 
& $\mathcal{E},\;\mathcal{E}_{\pm}$ 
& \qquad $[\mathcal{E}] = L$ \\

Electric charges (source and probe) 
& $Q,\; q$ 
& \qquad $[Q]=[q]=L$ \\

Charge-to-mass ratio 
& $q/m$ 
& \qquad $[q/m] = 1$ \\

Electromagnetic four-potential 
& $A_{\mu}$ 
& \qquad $[A_{\mu}] = 1$ \\

Axial angular momentum 
& $p_\varphi = m r^2 \dot{\varphi}$ 
& \qquad $[p_\varphi] = L^{2}$ \\

Radial velocity (proper time) 
& $\dot r = dr/d\tau$ 
& \qquad $[\dot r] = 1$ \\

Relativistic effective potential 
& $V_{\rm eff} = \mathcal{E}_+/m$ 
& \qquad $[V_{\rm eff}] = 1$ \\

Binding energy 
& $\mathcal{E}_{\rm bind} = V_{\rm eff}-1$ 
& \qquad $[\mathcal{E}_{\rm bind}] = 1$ \\

Radial epicyclic frequency (proper time) 
& $\omega_r$ 
& \qquad $[\omega_r] = L^{-1}$ \\

Radial epicyclic frequency (coordinate time) 
& $\Omega_r$ 
& \qquad $[\Omega_r] = L^{-1}$ \\

Periastron precession per orbit
& $\Delta\varphi$ 
& \qquad $[\Delta\varphi] = 1$ \\

Ricci scalar curvature 
& $R$ 
& \qquad $[R] = L^{-2}$ \\

Kretschmann invariant 
& $K = R_{\mu\nu\rho\sigma}R^{\mu\nu\rho\sigma}$ 
& \qquad $[K] = L^{-4}$ \\

\hline
\end{tabular}

\caption{\fontsize{8}{9}\selectfont
Dimensional analysis of geometric, electromagnetic, and dynamical quantities governing the motion of a charged test particle in the massless charged spacetime, formulated in geometrized units ($G=c=1$). In this convention, spacetime coordinates, proper time, mass, energy, and electric charge all have dimensions of length. Velocities, electromagnetic potentials, and normalized effective potentials are dimensionless, angular momentum scales as $[L]^2$, and curvature invariants ($R$ and $K$) carry inverse powers of length (see Appendix A).}
\label{tab:dims}
\end{table*}

\vspace{0.05cm}
\setlength{\parindent}{0pt}

In Table~\ref{tab:comparison}, we present a detailed comparison of analytical and numerical results for circular orbits of a charged particle ($m=1$, $q=-1$) around a central charge $Q=1$ in the weak-field regime, where the circular orbit radius satisfies $r_c \gg Q$. Analytical values are 
computed using the weak-field formulas derived in subsection~\ref{weak-field}, while numerical values are obtained by solving the full circular orbit condition including curvature corrections. The comparison includes circular orbit radii $r_c$, radial epicyclic frequencies $\omega_r$, azimuthal frequencies $\omega_\varphi$, and retrograde precession $\Delta \varphi$ for angular momenta $L=3$ to $L=10$. Relative deviations between analytical and numerical values illustrate the small effect of spacetime curvature in the weak-field regime. All precession values are negative, confirming retrograde motion.

\begin{table*}[ht]
\centering
\caption{\fontsize{8}{9}\selectfont Comparison of analytical and numerical results for circular orbits of a charged particle ($m=1$, $q=-1$) around a central charge $Q=1$ in the weak-field regime ($r_c \gg Q$). The table reports the circular orbit radius $r_c$, radial epicyclic frequency $\omega_r$, azimuthal frequency $\omega_\varphi$, and retrograde precession $\Delta \varphi$ for angular momenta $L=3,\dots,10$. Analytical values are obtained from the leading-order weak-field expressions in Sec.~\ref{weak-field}, while numerical results are computed by solving the full circular orbit condition $\mathcal{E}_+'(r_c) = 0$, including the first-order curvature corrections. All retrograde precession values are negative, indicating backward orbital precession.}
\vspace{2mm} 
\rowcolors{2}{gray!15}{white} 
\begin{tabular}{c c c c c c c c}
\toprule
$L$ & $r_c^{\rm analytic}$ & $r_c^{\rm numeric}$ & $\omega_r^{\rm analytic}$ & $\omega_r^{\rm numeric}$ & $\omega_\varphi^{\rm analytic}$ & $\omega_\varphi^{\rm numeric}$ & $\Delta \varphi$ (rad) \\
\midrule
3 & 9.0000 & 9.0020 & 0.037037 & 0.037030 & 0.036987 & 0.036980 & -0.3491 \\
4 & 16.0000 & 16.0020 & 0.015625 & 0.015624 & 0.015619 & 0.015618 & -0.1963 \\
5 & 25.0000 & 25.0010 & 0.008000 & 0.007999 & 0.007996 & 0.007995 & -0.1257 \\
6 & 36.0000 & 36.0005 & 0.0046296 & 0.004629 & 0.004627 & 0.004626 & -0.0873 \\
7 & 49.0000 & 49.0008 & 0.002915 & 0.002914 & 0.002913 & 0.002912 & -0.0641 \\
8 & 64.0000 & 64.0009 & 0.001953 & 0.001953 & 0.001952 & 0.001952 & -0.0491 \\
9 & 81.0000 & 81.0010 & 0.001371 & 0.001371 & 0.001371 & 0.001371 & -0.0387 \\
10 & 100.0000 & 100.0010 & 0.001000 & 0.001000 & 0.001000 & 0.001000 & -0.0314 \\
\bottomrule
\end{tabular}
\label{tab:comparison}
\end{table*}

\vspace{0.05cm}
\setlength{\parindent}{0pt}

The Figure~\ref{fig:pot-1} illustrates the dynamics of a charged particle with mass $m = 1$ and charge $q = -1$ orbiting a central Coulomb charge $|Q| = 2.4$ for angular momenta $L = 1, 2, 3$. The radial energy $\mathcal{E}_+(r)$ demonstrates the combined contributions of the particle's rest mass, Coulomb attraction, and angular momentum, with circular orbits $r_c$ identified as vertical dashed lines and the outermost radial barrier $r_{\ast}$ highlighted in purple. The effective potential $V_{\rm eff}(r)$ emphasizes the purely radial energy landscape, showing the locations of circular orbits relative to $r_{\ast}$. Radial oscillations $r(t)$ around these orbits are depicted with shaded envelopes representing the oscillation amplitude, demonstrating the stability of motion near $r_c$ while respecting the minimum radius $r_{\ast}$. Two-dimensional precessing orbits in the $xy$ plane reveal retrograde precession of periapsis due to the curvature term, with the orbit envelopes showing the maximal and minimal radial excursions and the outermost barrier $r_{\ast}$ clearly indicated. Together, these panels visualize how angular momentum and Coulomb interaction shape the particle’s motion and the retrograde shift of orbital trajectories.

\subsection{Strong-Field Dynamics and Orbital Stability}

\setlength{\parindent}{0pt}

In the strong-field limit, corresponding to $u \to \pi/2$ (equivalently $r \to r_{\ast}$), it is convenient to introduce a small expansion parameter:
\[
u = \frac{\pi}{2} - \epsilon, 
\qquad 0 < \epsilon \ll 1,
\] 
for which the trigonometric functions diverge as:
\[
\tan u = \cot\epsilon \;\simeq\; \frac{1}{\epsilon} - \frac{\epsilon}{3} + \mathcal{O}(\epsilon^3), 
\qquad 
\sec u = \csc\epsilon \;\simeq\; \frac{1}{\epsilon} + \frac{\epsilon}{6} + \mathcal{O}(\epsilon^3).
\] 
The future-directed energy branch then admits the expansion:
\begin{equation}
\mathcal{E}_+(u) \;\simeq\; \frac{q}{\epsilon} 
+ \sqrt{\;\frac{m^2}{\epsilon^2} 
+ \frac{L^2 (\pi/2)^2}{|Q|^2}\,\frac{1}{\epsilon^4}} \,,
\end{equation}
where we consider $q<0$ (particle) and $Q>0$ (background/source). For nonzero angular momentum ($L\neq 0$), the centrifugal term dominates, giving the leading scaling:
\[
\mathcal{E}_+(u) \;\sim\; \frac{L \pi}{2 |Q|} \,\frac{1}{\epsilon^2}.
\] 
For purely radial motion ($L=0$), the divergence is milder: 
\[
\mathcal{E}_+(u) \;\sim\; \frac{1}{\epsilon}.
\] 
This distinction shows that angular momentum strongly amplifies the confining barrier, while radial trajectories ($L=0$) approach it more gradually. 

\vspace{0.1cm}

The ability of a radial particle to approach the outermost shell at \(r_{\ast}\) depends on the charge-to-mass ratio $|q|/m$: typical values $|q|/m < 1$ enforce a turning point outside $r_{\ast}$, while larger ratios allow closer approach due to electrostatic attraction. Circular orbits must lie strictly outside the singular shell ($r>r_{\ast}$). The hypersurface $r = r_{\ast}$ acts as an impenetrable barrier: for $L \neq 0$, the centrifugal divergence ensures reflection before $r_{\ast}$; for $L=0$, accessibility is controlled by $|q|/m$ and the conserved energy. Radial dynamics are governed by the function $\mathcal{R}(r)$ defined in Eq.~\eqref{eq:Rdef-extended}, whose zeros specify turning points separating classically allowed and forbidden regions. In the strong-field regime, these zeros accumulate near $r_{\ast}$, producing either tightly confined oscillations or unstable equilibria. Orbital stability is quantified by the proper-time radial epicyclic frequency $\omega_r$ evaluated at the circular orbit radius $r_c$. The behavior of $\omega_r^2$ is determined by the curvature of the effective radial potential:
\begin{itemize}
\item \textbf{Stable orbits} ($\omega_r^2>0$): $\mathcal{R}''(r_c)<0$. Small radial perturbations lead to harmonic oscillations around $r_c$.  
\item \textbf{Marginal stability} ($\omega_r=0$): $\mathcal{R}''(r_c)=0$. The restoring force vanishes; the orbit sits at the edge of stability. This typically occurs for $L=0$ and $|q|/m \lesssim 1$, just outside $r_{\ast}$.  
\item \textbf{Instability} ($\omega_r^2<0$, $\omega_r$ imaginary): $\mathcal{R}''(r_c)>0$. Small radial perturbations grow exponentially. This arises for $L \neq 0$ or when the centrifugal or electrostatic terms create a steep potential slope near $r_{\ast}$.
\end{itemize}
Near $r_{\ast}$, the strong divergence of $\mathcal{E}_+(u)$ imposes a hard-wall confinement. For $L \neq 0$, turning points are pushed outward, producing narrow oscillatory regions; for $L=0$, the approach to $r_{\ast}$ is controlled by electrostatic attraction and gravitational curvature. Circular orbits near local maxima of $V_{\rm eff}(r)$ are generically unstable, and stable orbits cannot exist arbitrarily close to $r_{\ast}$.

\vspace{0.05cm}
\setlength{\parindent}{0pt}

The singular hypersurface at $r_{\ast}$ partitions the radial domain into isolated zones of motion, producing distinct families of bound and scattering states. This hard-wall confinement contrasts with black-hole dynamics, where horizons, rather than divergent shells, impose boundaries. The strong-field regime complements the weak-field description: at large radii, motion is approximately Keplerian with small retrograde precession, while near $r_{\ast}$, dynamics are dominated by the diverging effective potential. Together, these limits provide a continuous and unified picture of charged-particle motion across all accessible radial scales.

\vspace{0.05cm}
\setlength{\parindent}{0pt}

To further contextualize the present horizonless charged geometry, it is instructive to contrast it with well-known charged black-hole spacetimes, such as the RN \cite{RN} and Bardeen solutions \cite{B}. In the weak-field regime ($r \gg |Q|$), all three configurations exhibit Coulomb-like contributions to the effective potential and standard centrifugal barriers due to angular momentum. However, their global geometric and causal structures differ substantially. The RN metric \(f_{RN}(r) = 1 - \frac{2M}{r} + \frac{Q^2}{r^2}\) features two event horizons at \(r_\pm = M \pm \sqrt{M^2 - Q^2}\) \cite{RN} and a photon-sphere whose location is determined by \(M\) and \(Q\), enclosing a central curvature singularity. In contrast, the massless charged geometry \eqref{o-metric-red} is strictly horizonless, with curvature singularities occurring at discrete radial shells \(r_n = \frac{|Q|}{(n+1/2)\pi}, \, n = 0,1,2,\dots\), which act as dynamical barriers confining timelike particle motion to regions between successive shells, with the outermost shell at \(r_\ast = 2|Q|/\pi\) defining a classical boundary for orbits with nonzero angular momentum. Regular black-hole spacetimes such as the Bardeen solution \(f_B(r) = 1 - \frac{2 M r^2}{(r^2 + g^2)^{3/2}}\) eliminate central curvature singularities while retaining one or two horizons depending on the mass \(M\) and magnetic charge \(g\) \cite{B}. By contrast, the present massless charged configuration preserves singular shells yet remains horizonless, representing a physically distinct class of charge-dominated spacetimes. This setting provides a clean theoretical laboratory to probe curvature generated solely by electromagnetic fields, enabling detailed investigations of charge-curvature interactions, particle dynamics, and semiclassical properties without the confounding influence of mass-driven horizons.

\section{Mapping to a one-electron atom}\label{sec:mapping}

\setlength{\parindent}{0pt}

The dynamics of a charged particle on the background \eqref{o-metric-red} may be semiclassically mapped to a hydrogen-like one-electron system. This correspondence is valid in the regime where characteristic orbital radii satisfy \(r \gg |Q|\)(in units where $c=1=\hbar$), allowing metric functions such as \(\cos(|Q|/r)\), \(\sec(|Q|/r)\), and \(\tan(|Q|/r)\) to be systematically expanded in powers of \(|Q|/r\). Particle velocities are assumed nonrelativistic, with kinetic energies small compared to the rest mass energy \(m\) (in units where \(c=1\)), justifying a Schr\"odinger or semiclassical Bohr description. The particle is treated as a test particle, so its electromagnetic and gravitational backreaction is negligible. Finally, the quantum probability density should remain concentrated far from the outermost curvature singular shell \(r_\ast = 2|Q|/\pi\), ensuring rapid convergence of the perturbative expansion. In this controlled regime, the dominant dynamics is Coulombic, with curvature-induced corrections that are small and systematically computable, in principle.

\vspace{0.05cm}
\setlength{\parindent}{0pt}

Starting from the exact first integral for timelike charged motion, we denoted the positive-energy branch by \(\mathcal{E}_+(r)\). In the weak-field regime \(r \gg |Q|\), the expansion in \eqref{E-expansion} reads:
\begin{equation*}
\mathcal{E}_+(r) -m \simeq  \frac{q Q}{r} + \frac{L^2}{2 m r^2} + \frac{m Q^2}{2 r^2} + \mathcal{O}(r^{-3}) \,.
\end{equation*}
This form defines the effective potential for slow particles:
\begin{equation*}
V_{\rm eff}(r) \equiv \mathcal{E}_+(r) - m \simeq \frac{q Q}{r} + \frac{L^2}{2 m r^2} + \frac{m Q^2}{2 r^2} + \mathcal{O}(r^{-3}) \,,
\end{equation*}
where the leading term is Coulombic, the second is the centrifugal term, and the third is a geometric correction due to curvature. Higher-order terms modify the centrifugal structure with explicit \(|Q|/r\) dependence.

\vspace{0.05cm}
\setlength{\parindent}{0pt}

Within this approximation, one can map the system to hydrogenic variables as \(q \leftrightarrow -e\), \(Q \leftrightarrow Ze\), \(m \leftrightarrow m_e\), and \(L \leftrightarrow n \hbar\) semiclassically. The Coulomb term then becomes \(-Z e^2/r\), and the semiclassical orbital radius follows from balancing centrifugal and Coulomb forces,
\begin{equation}
r_c \simeq \frac{L^2}{m |q Q|} \,.
\end{equation}
With \(L = n \hbar\) and \(q Q = -e \cdot Ze\), this reproduces the Bohr-like radius \cite{L-3,LL-3}
\begin{equation}
a_n = \frac{n^2 \hbar^2}{m_e Z e^2} \,,
\end{equation}
which establishes the expected semiclassical hierarchy in planar geometry.

\vspace{0.05cm}
\setlength{\parindent}{0pt}

\begin{figure}[ht] 
\centering 
\includegraphics[scale=0.60]{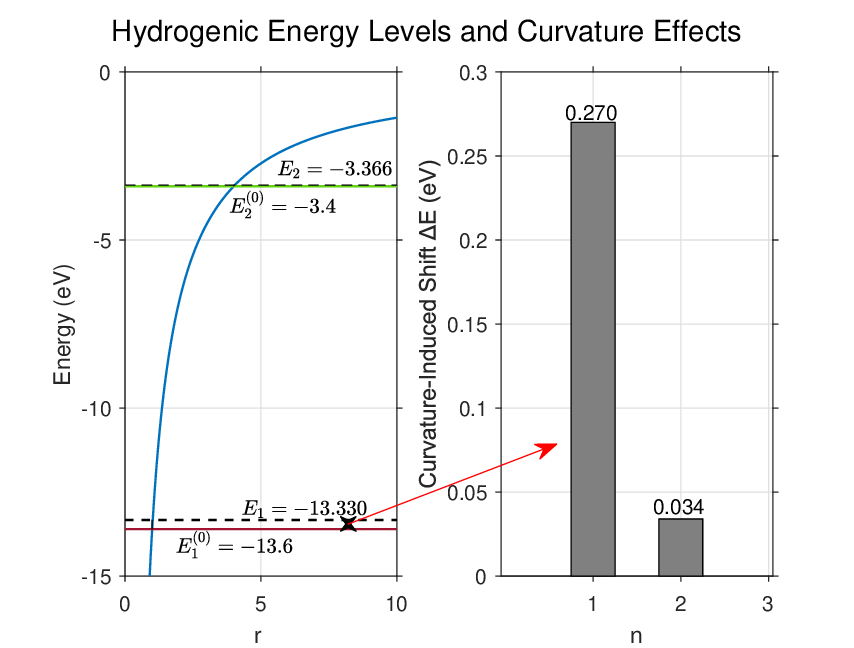} 
\caption{\fontsize{8}{9}\selectfont Hydrogenic energy levels (left) and curvature-induced shifts (right). The Coulomb potential is shown in blue, with unperturbed hydrogenic energies for $n = 1, 2$ depicted as solid lines. Curvature-perturbed energies are indicated by dashed black lines. The bar plot quantifies curvature-induced shifts, highlighting that the ground state ($n=1$) experiences the largest shift.} 
\label{fig:E-shift} 
\end{figure}

In the nonrelativistic quantum regime, the unperturbed Hamiltonian $H_0$ is
\begin{equation}
H_0 = \frac{\textbf{p}^2}{2m} + \frac{q Q}{r}.
\end{equation}
However, this form does not fully capture the influence of the curved spacetime \eqref{o-metric-red}. In the weak-field regime, the leading order geometric correction:
\begin{equation}
\delta V(r) = \frac{m Q^2}{2 r^2}
\end{equation}
can be treated perturbatively. To first order, the energy shift of a hydrogenic eigenstate \(|n \ell\rangle\) is \cite{L-3,LL-3}
\begin{equation}
\Delta E^{(1)}_{n\ell} = \langle n \ell | \delta V | n \ell \rangle = \frac{m Q^2}{2} \langle r^{-2} \rangle_{n \ell} \,,
\end{equation}
with \(\langle r^{-2} \rangle_{n\ell}\) finite for all \(\ell \ge 0\). The expectation values \(\langle r^{-2} \rangle_{n\ell}\) can be computed explicitly using 2+1 dimensional hydrogenic wavefunctions \cite{L-3,LL-3,AO-2024}, giving
\(\langle r^{-2} \rangle_{n\ell} = 1/(a_n^2 (\ell + 1/2))\) for \(\ell \ge 0\), consistent with standard planar quantum mechanics \cite{AO-2024}. The unperturbed binding energies $E^{(0)}_{n}$ are given by:
\begin{equation}
E^{(0)}_{n} = \mathcal{E}_n^{(0)} - m \simeq -\,\frac{m (q Q)^2}{2 \hbar^2 n^2} \,,
\end{equation}
which, for hydrogen (\(Z=1,\, Q=e\)), yield:
\begin{equation}
E_1^{(0)} \simeq -13.6~\mathrm{eV}, \qquad
E_2^{(0)} \simeq -3.40~\mathrm{eV}.
\end{equation}
Here, \(E^{(0)}_{n} \simeq -\,\frac{\mu e^4}{2 (4\pi \varepsilon_0)^2 \hbar^2 n^2}\), where the reduced mass is 
\(\mu \approx m_e \!\left( 1 - \frac{m_e}{m_p} \right)\), i.e., \(\mu = 9.104425 \times 10^{-31}\ \mathrm{kg}\) and \(\mu/m_e \approx 0.999455\) in SI units. The first-order curvature-induced corrections are
\begin{equation}
\Delta E_1^{(1)} \simeq 0.27~\mathrm{eV}, \quad
\Delta E_2^{(1)} \simeq 0.034~\mathrm{eV}.
\end{equation}
Hence, the total energies become:
\begin{equation}
E_n = E_n^{(0)} + \Delta E_n^{(1)} \simeq 
-13.33~\mathrm{eV}, \,-3.366~\mathrm{eV} \quad \text{for } n=1,2.
\end{equation}
These results confirm the validity of the perturbative approach (see also Figure~\ref{fig:E-shift}), since 
\(\Delta E_n^{(1)} \ll |E_n^{(0)} - E_{n+1}^{(0)}|\). Higher-order terms of order \(\mathcal{O}(r^{-3})\) are negligible for \(r \gg |Q|\), ensuring rapid convergence of the perturbative series \cite{PRA} (see Appendix B).

\vspace{0.05cm}
\setlength{\parindent}{0pt}

The classical radial epicyclic frequency, derived from the effective potential, satisfies \(\omega_r^2 \simeq \frac{\mathcal{E}_+''(r_c)}{m}\) in the weak-field limit, with curvature corrections entering at higher order in \(|Q|/r\). Explicitly, differentiating the expanded \(\mathcal{E}_+(r)\) gives \(\mathcal{E}_+''(r) = 2 q Q / r^3 + 3 L^2/(m r^4) + 3 m Q^2/r^4 + \mathcal{O}(r^{-5})\). Evaluated at \(r_c\), this reproduces the classical radial oscillation frequency, consistent with semiclassical hydrogenic predictions. The semiclassical radial oscillation spectrum thus agrees with the hydrogenic semiclassical treatment to leading order, validating the energy and radius identifications.

\vspace{0.05cm}
\setlength{\parindent}{0pt}

The hydrogen-like correspondence developed here is formally elegant but intrinsically approximate, valid primarily in the weak-field, large-radius regime where \(|Q|/r \ll 1\). Near the outermost curvature singular shell at \(r_\ast = 2|Q|/\pi\), several effects limit this analogy: wavefunction leakage, metric-dependent relativistic corrections, deviations of the exact gauge \(A_t = -\tan(|Q|/r)\) from the Coulomb potential \(-Q/r\), and potential issues with self-adjointness of the effective Hamiltonian. These factors define a natural boundary for the applicability of the mapping and indicate regions where a full relativistic treatment on the exact metric with consistent boundary conditions is required. Quantitatively, the correspondence holds when the typical orbital radius \(r_{\rm typ} \gg |Q|\), curvature-induced energy shifts remain small compared to interlevel spacings, the test-particle approximation is valid, and wavefunction support near \(r_\ast\) is negligible. In this regime, the effective potential can be expanded in powers of \(|Q|/r\) and treated perturbatively, with the leading-order Hamiltonian \(V_0(r) = q Q/r\) providing a systematically improvable approximation. The singular shell plays a role analogous to a nuclear core, strongly constraining short-distance quantum behavior, but unlike a smooth potential, it introduces metric-induced corrections affecting both kinetic and potential operators. To rigorously quantify the range of validity of this hydrogenic mapping, we introduce the (weak-field) expansion parameter \(u = |Q|/r\) and perform a controlled asymptotic expansion of the metric functions entering the effective Hamiltonian. Explicit error bounds are derived using Taylor expansions with Lagrange remainder \cite{book-1,book-2}, ensuring uniform convergence for \(r \gg |Q|\). The curvature-corrected hydrogenic energy shift is then given by \(\Delta E(r) = m Q^2/(2 r^2) + \mathcal{O}(|Q|^3/r^3)\), with the relative truncation error bounded by \(2 |Q|/r\). This rigorous derivation and error analysis are presented in detail in Appendix~B, providing a quantitative foundation for the controlled weak-field perturbative approach.

\section{Curvature-Corrected Thermodynamic Properties}
\label{sec:thermo}

The single-particle spectrum derived in Section~\ref{sec:mapping} can be incorporated into spectral-based thermodynamics by constructing the partition function over a controlled set of bound states. For definiteness, we restrict to the s-wave manifold and adopt unperturbed hydrogenic energies (in electronvolts):
\begin{equation}
E_n^{(0)}=-\frac{13.6}{n^2}, \qquad n=1,2,\dots,
\label{E0-def}
\end{equation}
augmented by curvature-induced perturbative shifts \(\Delta E_n^{(1)}\). Using the findings:
\[
\Delta E_1^{(1)} \simeq +0.270\ \mathrm{eV}, \qquad 
\Delta E_2^{(1)} \simeq +0.034\ \mathrm{eV},
\]
we consider a power-law interpolation of the shifts, yielding \(\Delta E_n^{(1)} \propto n^{-p}\) with \(p\approx 3\). Motivated by this observation and aiming for a minimal phenomenological description, we may adopt a simple model:
\begin{equation}
\Delta E_n^{(1)} = \frac{\Delta E_1^{(1)}}{n^3}, \qquad n \ge 1,
\label{dE-model}
\end{equation}
which reproduces the second-level shift \(\Delta E_2^{(1)} \simeq 0.0338\ \mathrm{eV}\). 
Here it should be emphasized that all thermodynamic quantities are defined operationally with respect to asymptotic observers located in the weak-field region ($r\gg |Q|$), where the metric varies slowly and Tolman’s redshift relation \(T(r)\sqrt{-g_{tt}(r)} = T_\infty, \quad g_{tt}(r) = -\cos^{-2}(|Q|/r)\), is valid \cite{Tolman,Tolman-2}. No assumption of global thermal equilibrium across the sequence of curvature singular shells is made. Curvature effects are parametrically strongest for the lowest-lying bound states; accordingly, the curvature-induced energy shifts are evaluated for the two deepest levels ($n=1,2$). The used power-law index $p\simeq 3$ is therefore to be interpreted strictly as a low-$n$ effective exponent $p_{\rm eff}$ characterizing the non-asymptotic part of the spectrum.
Accordingly, the total spectrum entering canonical sums is thus:
\begin{equation}
E_n = E_n^{(0)} + \Delta E_n^{(1)}.
\label{En-total}
\end{equation}
The practical calculation requires truncating the Rydberg series at a finite integer \(n_{\max}\). This truncation reflects the system's finite spatial extent, screening effects, or breakdown of the test-particle approximation; convergence must therefore be checked by varying \(n_{\max}\). With \(\beta \equiv 1/(k_{\mathrm{B}}T)\) and energies in eV (so \(k_{\mathrm{B}} = 8.617333262145\times 10^{-5}\ \mathrm{eV/K}\)), the canonical partition function reads \cite{Landau-Lifshitz,QSM-1,QSM-2,QSM-3}:
\begin{equation}
Z(\beta) = \sum_{n=1}^{n_{\max}} g_n\, e^{-\beta E_n},
\label{Z-def}
\end{equation}
where \(g_n = 1\) for the s-wave truncation, and canonical occupation probabilities are \cite{Landau-Lifshitz,QSM-1,QSM-2,QSM-3}:
\begin{equation}
p_n(\beta) = \frac{e^{-\beta E_n}}{Z(\beta)}.
\label{pn-def}
\end{equation}
Thermodynamic potentials are obtained in the standard way \cite{Landau-Lifshitz}:
\begin{align}
F(\beta) &= -\frac{1}{\beta} \ln Z(\beta),\\
U(\beta) &= \sum_{n=1}^{n_{\max}} p_n(\beta)\, E_n = -\frac{\partial \ln Z}{\partial \beta},\\
S(\beta) &= \frac{U(\beta) - F(\beta)}{T},\\
C_V(\beta) &= k_{\mathrm{B}} \beta^2 \Big( \langle E^2 \rangle - \langle E \rangle^2 \Big),
\end{align}
with \(\langle X \rangle \equiv \sum_n p_n X_n\). Here, \(F\) is the Helmholtz free energy, \(U\) is the internal energy, \(S\) is the entropy, and \(C_V\) is the heat capacity at constant volume. Moreover, the identity \(F = U - T S\) serves as a stringent numerical consistency check. All numerical values can be obtained via a stable direct evaluation of the truncated sums. To avoid overflow/underflow in exponentials, we employ the log-sum-exp technique: for a given set of energies \(\{E_n\}_{n=1}^{n_{\max}}\), we define \(E_{\min} = \min_n E_n\) and shifted weights \(\tilde z_n = \exp[-\beta(E_n - E_{\min})]\). The partition function is then \(Z = \tilde Z \exp(-\beta E_{\min})\) with \(\tilde Z = \sum_n \tilde z_n\), and normalized probabilities are \(p_n = \tilde z_n / \tilde Z\). Thermodynamic quantities follow as:
\begin{equation}
\begin{split}
&F = -\beta^{-1}(\ln \tilde Z - \beta E_{\min}), \quad U = \sum_n p_n E_n, \\
&S = \frac{U - F}{T}, \quad C_V = k_{\mathrm{B}} \beta^2 \left(\sum_n p_n E_n^2 - U^2\right). 
\end{split}  
\end{equation}
The same routine applies seamlessly to both the unperturbed and curvature-corrected spectra, with the resulting curvature-induced shifts, $\Delta X = X - X^{(0)}$, evaluated directly. Numerical verification must obey that $F = U - T S$ \cite{Landau-Lifshitz}.

\vspace{0.05cm}
\setlength{\parindent}{0pt}

For small curvature corrections, it is instructive to expand to first order in \(\Delta E_n^{(1)}\). Defining the unperturbed partition function and probabilities:
\[
Z^{(0)}(\beta) = \sum_{n=1}^{n_{\max}} e^{-\beta E_n^{(0)}}, \qquad
p_n^{(0)}(\beta) = \frac{e^{-\beta E_n^{(0)}}}{Z^{(0)}(\beta)},
\]
one finds to linear order:
\begin{align}
\Delta F &\simeq \langle \Delta E^{(1)} \rangle_0, \\
\Delta U &\simeq \langle \Delta E^{(1)} \rangle_0 - \beta \Big( \langle E^{(0)} \Delta E^{(1)} \rangle_0 - \langle E^{(0)} \rangle_0 \langle \Delta E^{(1)} \rangle_0 \Big), \\
\Delta S &\simeq - k_{\mathrm{B}} \beta^2 \Big( \langle E^{(0)} \Delta E^{(1)} \rangle_0 - \langle E^{(0)} \rangle_0 \langle \Delta E^{(1)} \rangle_0 \Big),
\end{align}
while \(C_V\) is computed directly from the variance definition for numerical stability. Convergence with respect to \(n_{\max}\) must be carefully tested. Representative results are summarized in Table~\ref{tab:convergence}.
\begin{table}[h!]
\centering
\fontsize{8}{9}\selectfont
\setlength{\tabcolsep}{3pt}
\caption{\fontsize{8}{9}\selectfont Convergence test of curvature-induced free-energy shifts $\Delta F$ [eV] and occupation probabilities at different truncation levels $n_{\max}$ and temperatures $T$. Values computed using a numerically stable log-sum-exp evaluation.}
\label{tab:convergence}
\vspace{0.3cm}
\begin{tabular}{c c c c c}
\toprule
$T$ [K] & $n_{\max}$ & $\Delta F$ [eV] & $p_1$ & $p_{n_{\max}}$ \\
\midrule
$300$        & $100$ & $0.27000$ & $0.9999999999999$ & $1.23\times10^{-224}$ \\
$300$        & $200$ & $0.27000$ & $0.9999999999999$ & $1.18\times10^{-224}$ \\
$300$        & $300$ & $0.27000$ & $0.9999999999999$ & $1.17\times10^{-224}$ \\
$10^{4}$     & $100$ & $0.26999$ & $0.999970$ & $1.92\times10^{-7}$ \\
$10^{4}$     & $200$ & $0.26999$ & $0.999951$ & $1.91\times10^{-7}$ \\
$10^{4}$     & $300$ & $0.26998$ & $0.999931$ & $1.91\times10^{-7}$ \\
$2\times10^{4}$ & $100$ & $0.25870$ & $0.954539$ & $4.18\times10^{-4}$ \\
$2\times10^{4}$ & $200$ & $0.24904$ & $0.916259$ & $4.01\times10^{-4}$ \\
$2\times10^{4}$ & $300$ & $0.24008$ & $0.880940$ & $3.85\times10^{-4}$ \\
\bottomrule
\end{tabular}
\end{table}
Representative curvature-induced shifts of thermodynamic quantities are reported in Table~\ref{tab:thermo-values}. At room temperature, the ensemble is essentially confined to the ground state, so free and internal energies coincide with the curvature-shifted ground-state value:
\[
F^{(0)} \simeq -13.600\ \mathrm{eV}, \qquad F \simeq -13.330\ \mathrm{eV},
\]
while entropy and heat capacity vanish. At higher temperatures, thermal occupation of excited states produces finite curvature-induced corrections. Free and internal energies are directly influenced by the mean level correction, whereas entropy and heat capacity reflect the redistribution of populations among excited states.
\begin{table}[h!]
\centering
\fontsize{8}{9}\selectfont
\setlength{\tabcolsep}{3pt}
\caption{\fontsize{8}{9}\selectfont Curvature-induced shifts of canonical thermodynamic quantities at representative temperatures. Values computed with $n_{\max}=300$.}
\label{tab:thermo-values}
\vspace{0.3cm}
\begin{tabular}{c c c c c}
\toprule
$T$ [K] & $\Delta F$ [eV] & $\Delta U$ [eV] & $\Delta S$ [$10^{-8}$ eV/K] & $\Delta C_V$ [$10^{-7}$ eV/K] \\
\midrule
$300$      & $+0.27000$  & $+0.27000$  & $0.00$   & $0.00$ \\
$10^{4}$   & $+0.26998$  & $+0.27022$  & $2.36$   & $3.11$ \\
\bottomrule
\end{tabular}
\end{table}

\begin{figure*}[ht]
\centering
\includegraphics[scale=0.50]{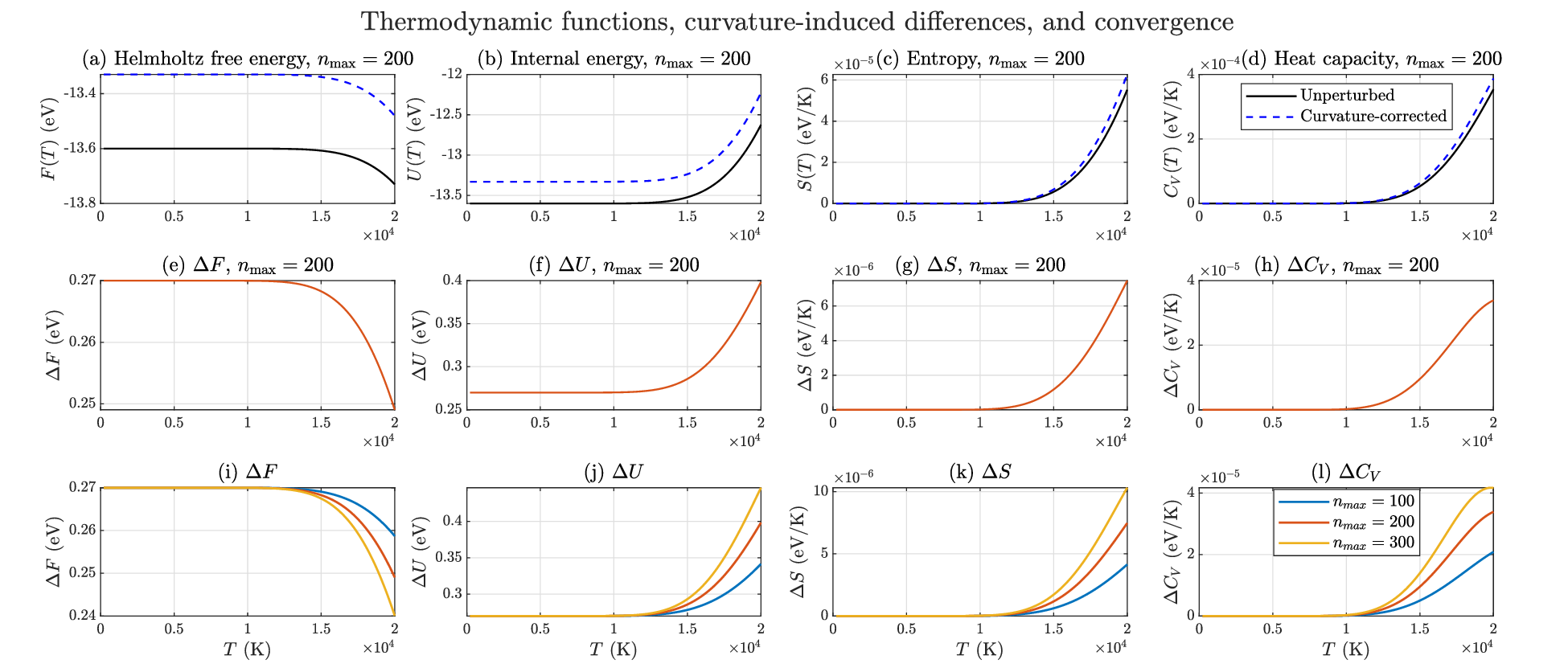}
\caption{\fontsize{8}{9}\selectfont Thermodynamic properties of the truncated hydrogenic spectrum with curvature corrections. Subplots (a--d) display the absolute canonical quantities: Helmholtz free energy $F(T)$, internal energy $U(T)$, entropy $S(T)$, and heat capacity $C_V(T)$ for $n_{\max}=200$, with solid black lines for the unperturbed energies $E_n^{(0)}$ and dashed blue lines including curvature shifts $\Delta E_n^{(1)}$. Subplots (e--h) present the curvature-induced differences $\Delta F$, $\Delta U$, $\Delta S$, and $\Delta C_V$. Subplots (i--l) show convergence for $n_{\max}=100,200,300$, illustrating the stability of canonical sums. Residuals $F-(U-TS)$ are smaller than $10^{-14}\,\mathrm{eV}$, confirming numerical consistency. All quantities are in eV or eV/K; the temperature axis is logarithmic to emphasize low- and high-temperature regimes.}
\label{fig:thermo-curvature}
\end{figure*}

\vspace{0.05cm}
\setlength{\parindent}{0pt}

Here, it is important to underline that the entropy $S$ and heat capacity $C_V$ computed here are effective semiclassical thermodynamic variables derived from curvature-corrected bound states. Because the present geometry is horizonless, these quantities do not represent Bekenstein-Hawking entropy and are not associated with Hawking radiation. Instead, they quantify the statistical response of curvature-induced energy levels to temperature variations.

\section{Summary and Discussion}
\label{sec:discussions}

\setlength{\parindent}{0pt}

We have performed a detailed analysis of the dynamics of charged test particles in a static, spherically symmetric spacetime sourced solely by an electric charge $Q$. This corresponds to the massless limit of a charged wormhole solution in the EMS system. The geometry, described by the metric in Eq.~\eqref{o-metric-red}, contains an infinite series of concentric curvature-singularity shells given in Eq.~\eqref{sing-points}. The outermost shell at $r_{\ast} = 2|Q|/\pi$ defines a true geometric boundary. For particles with nonzero angular momentum ($L \neq 0$), this shell acts as an impenetrable barrier. For purely radial motion ($L = 0$), accessibility depends on the charge-to-mass ratio $|q|/m$, with turning points occurring outside $r_{\ast}$ for particles approaching from infinity. The radial domain is thus divided into separate regions, forming a confinement structure reminiscent of classical potential walls.

\vspace{0.05cm}
\setlength{\parindent}{0pt}

Using the Lagrangian, we obtained exact first integrals for the temporal, azimuthal, and radial motion. The dynamics is governed by two energy branches, $\mathcal{E}_{\pm}(r)$, with the future-directed branch $\mathcal{E}_{+}(r)$ describing physical trajectories. The effective potential, expressed relative to the particle rest mass $m$, includes contributions from both the Coulomb interaction and spacetime curvature. In the weak-field regime ($r \gg |Q|$), the potential reduces to the Coulomb form with a centrifugal term and small curvature correction. These correction induces a retrograde periapsis precession,
\begin{equation}
\Delta\varphi \simeq -\frac{\pi m^{2}Q^{2}}{L^{2}},
\end{equation}
where the negative sign indicates a backward shift compared to the Newtonian case. For attractive Coulomb interactions ($qQ<0$), stable circular orbits exist at:
\begin{equation}
r_{c} = \frac{L^{2}}{m|qQ|},
\end{equation}
to leading order, and the radial epicyclic frequency is $\omega_{r}^{2}\simeq m^{2}|qQ|^{2}/L^{6}$. Increasing Coulomb coupling strengthens binding, while larger angular momentum lowers the oscillation frequency, reflecting the classical balance between central attraction and centrifugal repulsion.

\vspace{0.05cm}
\setlength{\parindent}{0pt}

Near $r_{\ast}$, the effective potential diverges. Introducing $\epsilon=\tfrac{\pi}{2}-|Q|/r$, one finds $\mathcal{E}_{+} \sim \epsilon^{-1}$ for radial motion and $\mathcal{E}_{+} \sim \epsilon^{-2}$ for nonradial motion. This divergence acts as a hard-wall barrier, which becomes more restrictive with increasing angular momentum. For $|q|/m<1$, the barrier is softened for purely radial trajectories, while nonradial motion remains strictly excluded. This establishes a hierarchy of confinement strengths, comparable to hard-wall models familiar from quantum mechanics.

\vspace{0.05cm}
\setlength{\parindent}{0pt}

At sufficiently large radii (see Appendix B), the system can be mapped to a hydrogen-like system. The Coulomb term dominates the potential, the centrifugal term balances orbital motion, and curvature corrections can be treated perturbatively. Using the semiclassical correspondence $q \leftrightarrow -e$, $Q \leftrightarrow Ze$, $L \leftrightarrow n\hbar$, and $m \leftrightarrow m_e$, the outermost singular shell $r_{\ast}$ plays a role analogous to the atomic nucleus, providing a short-distance boundary. The semiclassical orbital radii $a_n \sim n^2 \hbar^2 / (m |qQ|)$ reproduce the Bohr scaling, while the curvature-induced $r^{-2}$ term yields small, systematically computable energy shifts $\Delta E^{(1)}$. This analogy is quantitatively reliable when the wavefunction is localized far from $r_{\ast}$ and perturbative corrections remain small compared to interlevel spacing. The mapping thus provides a controlled connection between weak-field Coulombic orbits and the strong-field confinement induced by the singular shell. The system exhibits two complementary regimes. At large radii, particle motion resembles classical Coulomb or Keplerian dynamics with minor curvature corrections. Close to the outermost singular shell, motion is dominated by a steeply rising potential barrier that enforces strong spatial confinement. This framework provides a continuous description linking weak-field orbits to highly constrained dynamics near the singular shell, connecting classical orbital mechanics with exotic singular geometries.

\vspace{0.05cm}
\setlength{\parindent}{0pt}

Beyond the classical and semiclassical particle dynamics, curvature-induced corrections to the effective potential have direct consequences for the spectral thermodynamics of bound states. Constructing the partition function over s-wave bound states with energy shifts $\Delta E_n^{(1)}$ shows that curvature systematically increases the free and internal energies, weakens binding, and enhances thermal ionization. These thermodynamic effects become significant at temperatures comparable to the energy scale of the lowest bound-state corrections, whereas at low temperatures the ensemble remains effectively confined to the ground state. Entropy and heat capacity are altered subtly through correlations between unperturbed energies and curvature-induced shifts, providing a precise quantitative description of how the geometry modifies statistical properties. Integrating the results from classical particle dynamics, semiclassical mapping, and curvature-corrected thermodynamics establishes a consistent framework that links microscopic motion with macroscopic statistical behavior, demonstrating that the singular shell not only enforces spatial confinement but also produces measurable (in principle) shifts in the thermal characteristics of the system.

\vspace{0.05cm}
\setlength{\parindent}{0pt}

The results establish a clear and analytically tractable framework for charged-particle motion in horizonless, singular charged spacetimes. The combination of integrability, smooth connection to Coulomb dynamics at large radii, and hard-wall confinement near the singular shell demonstrates the value of this system as a theoretical laboratory for studying charged matter in geometries determined entirely by electromagnetic fields. The dynamical structures identified in this work may generate observable signatures, including anomalous perihelion precession of test-particle orbits and characteristic modifications of strong gravitational-lensing patterns. The absence of an event horizon gives rise to qualitatively distinct lensing phenomenology relative to black-hole spacetimes, potentially allowing discrimination through future high-precision astrophysical observations and controlled gravity-analog experimental platforms. Several extensions are suggested by this framework. Studying null geodesics could reveal the causal and optical properties of the singular shells, potentially producing distinctive lensing effects. A detailed analysis of radial and azimuthal oscillation frequencies would relate the results to classical celestial mechanics. Incorporating electromagnetic self-force or radiation-reaction effects could extend the model to dissipative systems. Semiclassical studies of wave propagation or quantized bound states may highlight confinement effects similar to a particle-in-a-box model. Finally, exploring rotational or perturbed versions of the geometry would test whether the confinement mechanisms persist in less symmetric conditions.







\begin{acknowledgments}
The authors thank Prof. M. Halilsoy for valuable and insightful discussions. The authors thank the anonymous referees and the editor for their constructive comments. H. H. thanks the Excellence Project FoS UHK 2203/2025--2026 for financial support.
\end{acknowledgments}

\appendix
\section*{Appendix A: Curvature Invariants of the Charged Geometry}
\label{app:curvature}

The spacetime under consideration is defined by the line element \eqref{o-metric-red}, in which \(Q\) denotes a real constant with dimensions of length. For compactness of notation one defines:
\begin{equation}
y(r)=\frac{Q}{r}, \qquad \Omega(r)=\cos y(r), \qquad S(r)=\sin y(r).
\end{equation}
The nonvanishing components of the metric tensor are therefore:
\[
g_{tt}=-\Omega^{-2},\qquad g_{rr}=\Omega^{2},\qquad g_{\theta\theta}=\Omega^{2}r^{2},\qquad
g_{\phi\phi}=\Omega^{2}r^{2}\sin^{2}\theta,
\]
and the inverse metric components are
\[
g^{tt}=-\Omega^{2},\qquad g^{rr}=\Omega^{-2},\qquad
g^{\theta\theta}=\frac{1}{\Omega^{2}r^{2}},\qquad
g^{\phi\phi}=\frac{1}{\Omega^{2}r^{2}\sin^{2}\theta}.
\]
The derivatives of the conformal factor are computed exactly. Since
\[
y'(r)=\frac{d}{dr}\Big(\frac{Q}{r}\Big)=-\frac{Q}{r^{2}}=-\frac{y}{r},
\]
one obtains:
\begin{equation}
\Omega'(r)=\frac{d}{dr}\cos y=-\sin y\, y'=\frac{y}{r}S,
\end{equation}
and
\begin{equation}
\Omega''(r)=\frac{d}{dr}\Big(\frac{y}{r}S\Big)
=-\frac{2y}{r^{2}}S-\frac{y^{2}}{r^{2}}\Omega.
\end{equation}
All partial derivatives of the metric then follow exactly:
\[
\partial_{r}g_{tt}=2\Omega^{-3}\Omega',\qquad
\partial_{r}g_{rr}=2\Omega\Omega',\qquad
\partial_{r}g_{\theta\theta}=2\Omega\Omega'r^{2}+2\Omega^{2}r,
\]
\[
\partial_{r}g_{\phi\phi}=(2\Omega\Omega'r^{2}+2\Omega^{2}r)\sin^{2}\theta.
\]
The Christoffel symbols are calculated directly from:
\[
\Gamma^{\rho}{}_{\mu\nu}=\frac12 g^{\rho\sigma}
\left(\partial_{\mu}g_{\nu\sigma}
+\partial_{\nu}g_{\mu\sigma}
-\partial_{\sigma}g_{\mu\nu}\right),
\]
which yields the following exact nonzero independent components:
\[
\Gamma^{t}{}_{tr}=\Gamma^{t}{}_{rt}=-\frac{\Omega'}{\Omega},\qquad
\Gamma^{r}{}_{tt}=-\Omega^{-5}\Omega',\qquad
\Gamma^{r}{}_{rr}=\frac{\Omega'}{\Omega},
\]
\[
\Gamma^{r}{}_{\theta\theta}=-r^{2}\frac{\Omega'}{\Omega}-r,\qquad
\Gamma^{r}{}_{\phi\phi}=-\Big(r^{2}\frac{\Omega'}{\Omega}+r\Big)\sin^{2}\theta,
\]
\[
\Gamma^{\theta}{}_{r\theta}=\Gamma^{\phi}{}_{r\phi}=\frac{\Omega'}{\Omega}+\frac{1}{r},\qquad
\Gamma^{\theta}{}_{\phi\phi}=-\sin\theta\cos\theta,\qquad
\Gamma^{\phi}{}_{\theta\phi}=\cot\theta.
\]
From these connection coefficients the Riemann tensor is constructed using:
\[
R^{\rho}{}_{\sigma\mu\nu}
=
\partial_{\mu}\Gamma^{\rho}{}_{\nu\sigma}
-
\partial_{\nu}\Gamma^{\rho}{}_{\mu\sigma}
+
\Gamma^{\rho}{}_{\mu\lambda}\Gamma^{\lambda}{}_{\nu\sigma}
-
\Gamma^{\rho}{}_{\nu\lambda}\Gamma^{\lambda}{}_{\mu\sigma}.
\]
The independent covariant components are obtained without approximation as:
\[
R_{trtr}=-\frac{\Omega''}{\Omega^{3}}+\frac{3\Omega'^{2}}{\Omega^{4}},
\]
\[
R_{t\theta t\theta}
=
-\frac{\Omega'}{\Omega^{3}}
\Big(r^{2}\frac{\Omega'}{\Omega}+r\Big),
\]
\[
R_{\theta r\theta r}
=-\Omega\, \Omega''\,r^2+\Omega''r^2-\Omega\, \Omega'\,r
\]
and
\[
 R_{\theta\phi\theta\phi}
=-\Omega^2\,r^2\,\sin^{2}\theta\Big(\frac{2r\Omega'}{\Omega}+\frac{r^2\Omega'^2}{\Omega^2}\Big)
\]
The Ricci tensor follows from the contraction \(R_{\mu\nu}=R^{\rho}{}_{\mu\rho\nu}\) and yields the exact results:
\[
R_{tt}=\frac{r\Omega'^2-\Omega(2\Omega'+r\Omega'')}{r\Omega^6} \qquad
R_{rr}=-\frac{\Omega''}{\Omega}-\frac{\Omega'^{2}}{\Omega^{2}}-\frac{2\Omega'}{\Omega r},
\]
\[
R_{\theta\theta}
=\frac{r^2\Omega'^2}{\Omega^2}-\frac{2r\Omega'}{\Omega}-\frac{r^2\Omega''}{\Omega},
\qquad
R_{\phi\phi}=\sin^{2}\theta\,R_{\theta\theta}.
\]
The Ricci scalar is obtained by the exact contraction \(R=g^{\mu\nu}R_{\mu\nu}\). Substituting the inverse metric components and simplifying only by means of exact algebraic and trigonometric identities, in particular \(\sin^{2}y+\cos^{2}y=1\), one arrives at:
\begin{equation}
R
=
\frac{2Q^{2}}{r^{4}\cos^{2}\!\left(\frac{Q}{r}\right)},
\end{equation}
with no series expansion or approximation of any kind. The Kretschmann scalar is constructed from the full contraction of the Riemann tensor,
\[
K=R_{\alpha\beta\gamma\delta}R^{\alpha\beta\gamma\delta},
\]
and after exact index raising and complete algebraic simplification gives:
\begin{widetext}
\begin{equation*}
\begin{aligned}
K(r) = &
\;\frac{4 Q^{2}}{r^{8}} \,
\sec^{8}\!\left(\frac{Q}{r}\right)
\Bigg[14 Q^{2} \sin^{4}\!\left(\frac{Q}{r}\right)- 12 r Q \, \sin^{2}\!\left(\frac{Q}{r}\right) 
       \sin\!\left(\frac{2Q}{r}\right)+ \left(\tfrac{5}{2} Q^{2} + 3 r^{2}\right)
   \sin^{2}\!\left(\frac{2Q}{r}\right)+ 48 Q r \, \cos^{2}\!\left(\frac{Q}{r}\right)
     \sin\!\left(\frac{2Q}{r}\right)+ 3 Q^{2} \cos^{4}\!\left(\frac{Q}{r}\right)
\Bigg].
\end{aligned}
\end{equation*}
\end{widetext}
These scalar invariants are fully coordinate independent and therefore provide an unambiguous characterization of the intrinsic curvature of the spacetime. The Ricci scalar diverges as \(r^{-4}\) when \(r\to r_n\) and diverges whenever the conformal factor vanishes, i.e. whenever \(\cos\left(\frac{Q}{r}\right)=0
\quad \Longleftrightarrow \quad
\frac{Q}{r}=\frac{\pi}{2}+n\pi,\quad n\in\mathbb{Z}\). The Kretschmann scalar exhibits divergences that precisely coincide with those of the Ricci scalar, scaling as \(r^{-8}\) near the origin and diverging at the discrete radii \(r_n\) where the conformal factor \(\cos(|Q|/r)\) vanishes. Since the Kretschmann scalar is a full contraction of the Riemann tensor, it is invariant under arbitrary smooth coordinate transformations. Consequently, these divergences are true curvature singularities and cannot be removed by any coordinate redefinition. Physically, the singularity at \(r=0\) represents a central curvature singularity, while each radius \(r_n\) defines a singular shell, corresponding to a hypersurface where the curvature invariants diverge (see Figure \ref{fig:curv-inv}). These shells act as dynamical barriers for timelike test particles, as the diverging tidal forces prevent penetration.  The discrete sequence of singular hypersurfaces represents a fundamentally different geometric structure compared to standard black hole spacetimes. Unlike a single central singularity, here the spacetime is punctuated by an infinite series of singular shells, each producing a breakdown of the classical geometric description. Approaching any shell, the curvature invariants diverge, indicating that classical general relativity ceases to provide a valid description, and quantum gravitational effects would presumably become dominant. Moreover, these results clarify the massless limit of the spacetime: as \(M \to 0\), the solution reorganizes non-analytically into the trigonometric form of the metric, generating the discrete shells. This confirms that the \(M\to 0\) limit is a singular perturbation, with the resulting geometry lying on a distinct, nonperturbative branch of the solution space. Therefore, the singular shells are intrinsic geometric features of the massless spacetime, with real physical consequences for geodesic motion and tidal effects.

\section*{Appendix B: Weak-Field Expansion and Error Bounds}\label{app:weak-field}

To rigorously justify the hydrogenic mapping, we introduce the dimensionless expansion parameter:
\begin{equation}
u = \frac{|Q|}{r},
\end{equation}
which measures the relative strength of spacetime curvature sourced by the charge. The perturbative expansion is explicitly restricted to the weak-curvature domain:
\begin{equation}
u \ll 1 \quad \Longleftrightarrow \quad r \gg |Q|,
\end{equation}
while the strong-field region \(r \sim |Q|\) is excluded from the series and treated via exact geometric relations. Using Taylor expansions with Lagrange remainder \cite{book-1,book-2}, the metric functions entering the effective Hamiltonian admit the controlled expansions \cite{book-1,book-2}:
\begin{align}
\sec^2 u &= 1 + u^2 + \frac{2}{3}u^4 + R_{\sec}(u),\\
\cos^2 u &= 1 - u^2 + \frac{1}{3}u^4 + R_{\cos}(u),
\end{align}
with remainder terms satisfying the explicit pointwise bounds \cite{book-1,book-2}:
\begin{equation}
|R_{\sec}(u)| \le \frac{2}{15}\,u^6 \sec^6 \xi, \quad
|R_{\cos}(u)| \le \frac{1}{45}\,u^6, \quad 0 < \xi < u.
\end{equation}
Substituting these expansions into the exact effective energy of a charged test particle and retaining terms up to order \(u^2\) yields the curvature-corrected hydrogenic shift:
\begin{equation}
\Delta E(r) = \frac{m Q^2}{2 r^2} + \mathcal{O}\Big(\frac{|Q|^3}{r^3}\Big),
\end{equation}
with the neglected higher-order contribution satisfying the uniform bound:
\begin{equation}
\Bigl| \Delta E(r) - \frac{m Q^2}{2 r^2} \Bigr| \le \frac{m |Q|^3}{r^3}, \quad r \gg |Q|.
\end{equation}
Consequently, the relative truncation error is explicitly bounded by:
\begin{equation}
\frac{\bigl|\Delta E(r) - \frac{m Q^2}{2 r^2}\bigr|}{\frac{m Q^2}{2 r^2}} \le 2 \frac{|Q|}{r}, \quad r \gg |Q|,
\end{equation}
which decays linearly with \(u\) in the weak-field limit. This controlled expansion establishes that the hydrogenic mapping is quantitatively reliable only for \(r \gg |Q|\), while the perturbative description necessarily breaks down as \(r \to |Q|\). The use of Lagrange remainder provides rigorous bounds for the truncation error, ensuring uniform convergence in all subsequent weak-field calculations \cite{book-1,book-2}.

\end{document}